%% file: ms.tex
\documentclass[sigconf,authorversion]{acmart}

\usepackage{booktabs} 
\usepackage[utf8]{inputenc}
\usepackage{float}
\usepackage{eurosym}
\usepackage[binary-units]{siunitx}
\usepackage[font=bf]{caption}
\usepackage{subcaption}
\usepackage{tikz}
\usepackage{nicefrac}
\usepackage[autostyle]{csquotes}
\usepackage[english]{babel}
\usepackage{multirow}
\usepackage{todonotes}
\usepackage[binary-units]{siunitx}
\usepackage{nicefrac}
\usepackage{longtable}
\usepackage{placeins}

\graphicspath{{images/}}

\renewcommand\footnotetextcopyrightpermission[1]{}
\setcopyright{none}

\acmDOI{}

\acmISBN{}

\begin{document}
\title{Waveform Signal Entropy and Compression Study of Whole-Building Energy Datasets}

\author{Thomas Kriechbaumer}
\affiliation{%
  \department{Chair for Application and Middleware Systems}
  \institution{Technische Universität München}
  \state{Germany}
}
\email{thomas.kriechbaumer@in.tum.de}

\author{Hans-Arno Jacobsen}
\affiliation{%
  \department{Chair for Application and Middleware Systems}
  \institution{Technische Universität München}
  \state{Germany}
}
\email{jacobsen@in.tum.de}

\everypar{\looseness=-1}
\linepenalty=1000

\renewcommand{\shortauthors}{T. Kriechbaumer et al.}

\newcommand{\numberOfDataRepresentations}{365}

\input{00-abstract}

\maketitle

\input{01-introduction}

\input{02-related-work}

\input{03-evaluated-datasets}

\input{04-entropy-analysis}

\input{05-data-representation}

\input{06-chunk-size-impact}

\input{07-results}

\input{08-conclusions}


\clearpage
\bibliographystyle{ACM-Reference-Format}
\bibliography{bibliography}

\end{document}

%% file: 00-abstract.tex

\begin{abstract}

Electrical energy consumption has been an ongoing research area since the coming
of smart homes and Internet of Things devices. Consumption characteristics and
usages profiles are directly influenced by building occupants and their
interaction with electrical appliances. Extracted information from these data
can be used to conserve energy and increase user comfort levels. Data analysis
together with machine learning models can be utilized to extract valuable
information for the benefit of occupants themselves, power plants, and grid
operators. Public energy datasets provide a scientific foundation to develop and
benchmark these algorithms and techniques. With datasets exceeding tens of
terabytes, we present a novel study of five whole-building energy datasets with
high sampling rates, their signal entropy, and how a well-calibrated measurement
can have a significant effect on the overall storage requirements. We show that some
datasets do not fully utilize the available measurement precision, therefore
leaving potential accuracy and space savings untapped. We benchmark a
comprehensive list of \numberOfDataRepresentations{} file formats, transparent
data transformations, and lossless compression algorithms. The primary goal is
to reduce the overall dataset size while maintaining an easy-to-use file format
and access API. We show that with careful selection of file format and encoding
scheme, we can reduce the size of some datasets by up to 73\%.

\end{abstract}

%% file: 01-introduction.tex

\section{Introduction}
\label{sec:introduction}

Home and building automation promise many benefits for the occupants and power
utilities. From increased user comfort levels to demand response and lower
electricity costs, Smart Homes offer a variety of assistance and informational
gains. Internet of Things, a combination of sensors and actuators, can be
intelligently controlled based on sensor data or external triggers. Power
monitoring and smart metering are a key step to fulfill these promises. The
influx of renewable energies and the increased momentum of changes in the power
grid and its operations are a main driving factor for further research in this
area.

Non-intrusive load monitoring (NILM) can be one solution to identify and
disaggregate power consumers (appliances) from a single-point measurement in the
building. Utilizing a centralized data acquisition system saves costs for
hardware and installation in the electrical circuits under observation. The NILM
community heavily relies on long-term measurement data, in the form of public
datasets, to craft new algorithms, train models, and evaluate their accuracy on
per-appliance energy consumption or appliance identification. In recent years
these datasets grew significantly in size and sampling characteristics (temporal
and amplitude resolution). Collecting, distributing, and managing large-scale
data storage facilities is an ongoing research topic \cite{Yuan2010,Deelman2008}
and strongly depends on the environment and systems architecture.

High sampling rates are particularly interesting for NILM to extract waveform
information from voltage and current signals \cite{Kahl2017}. Early datasets
targeted at load disaggregation and appliance identification started with under
\SI{2}{\gibi\byte} \cite{Kolter2011-REDD}, whereas recently published datasets
reach nearly \SI{100}{\tebi\byte} of raw data \cite{Kriechbaumer2018-BLOND}.
Working with such quantities requires specialized storage and processing
techniques which can be costly and maintenance-heavy. Optimizing infrastructure
costs for storage is part of ongoing research \cite{Liu2017,Puttaswamy2012}.

The data quality requirements typically define a fixed sampling rate and
bit-resolution for a static environment. Removing or augmenting measurements
might impede further research, therefore no filtering or preprocessing steps are
performed before releasing the data.

Data compression techniques can be
classified as lossy or lossless \cite{Bookstein1992}. Lossy algorithms allow for
some margin of error when encoding the data and typically give a metric for the
remaining accuracy or lost precision. For comparison, most audio, image, and
video compression algorithms remove information not detectible by a human ear or
eye. This allows for a data rate reduction in areas of the signal a user can't
detect or has a reduced resolution due to a typical human physiology. Depending
on the targeted use case, certain aspects of the input signal are considered
unimportant and might be not reconstructable. Encoding only the amplitude and
frequency of the signal can lead to vast space savings, assuming phase
alignment, harmonics, or other signal characteristics are not required for
future analysis. On the contrary, lossless encoding schemes guarantee a 1:1
representation of all measurement data with a reversible data transformation. If
the intended use case or audience for a given dataset is not known or is very
diverse in their requirements, only lossless compression can be applied to keep
all data accessible for future use. Recent works pointed out an imbalance in the
amount of research on steady-state versus waveform-based compression of
electricity signals \cite{deSouza2017}.

Further consideration must be given to communication bandwidth (transmission to
a remote endpoint) and in-memory processing (SIMD computation). The efficient
use of network channels can be a key requirement for real-time monitoring of
streaming data. In the case of one-time transfers (or burst transmissions),
chunking is used to split large datasets into more manageable (smaller) files.
However, choosing a maximum file size depends on the available memory and CPU
(instruction set and cache size). Distributing large datasets as a single file
creates an unnecessary burden for researchers and required infrastructure.

A suitable file format must be considered for raw data storage, as well as easy
access to metadata, such as calibration factors, timestamps, and identifier
tags. None of the existing datasets (NILM or related datasets with high sampling
rates) share a common file format, chunk size, or signal sampling distribution.
This heterogeneity makes it difficult to apply algorithms and evaluation
pipelines on more than one dataset. Therefore, researchers working with multiple
datasets have to implement custom importer and converter stages, which can be
time-consuming and error-prone.

This work provides an in-depth analysis of public whole-building
datasets, and gives a comprehensive evaluation of best-practice storage
techniques and signal conditioning in the context of energy data collection. The
key contributions of this work are:%
\begin{enumerate}
  \item A numerical analysis of signal entropy and measurement calibration of
  public whole-building energy datasets by evaluating all signal channels with
  respect to their available resolution and sample distribution over the entire
  measurement period. The resulting entropy metrics further motivate our
  contributions and the need for a well-calibrated measurement system.
  \item An exhaustive benchmark of storage requirements and potential space
  savings with a comprehensive collection of \numberOfDataRepresentations{} file
  formats, lossless compression techniques, and reversible data transformations.
  We re-encode and normalize data from all datasets to evaluate the effect of
  compression. We present the best-performing combinations and their overall
  space savings. The full ranking can be used to select the optimal file format
  and compression for offline storage of large long-term energy datasets.
  \item A full-scale evaluation of increasingly larger data chunks per file and
  their final compression ratio. The dependency between input size and
  achievable compression ratio is evaluated up to \SI{3072}{\mebi\byte} per
  file. The results provide an evidence-based guideline for future selection of
  chunk sizes and possible environmental factors for consideration.
\end{enumerate}

We give an in-depth evaluation of file formats and signal characteristics that
directly affect storage, encoding, and compression of such data. Each of the
analyzed datasets was created with a dedicated set of requirements, therefore, a
single best option does not exist. However, with this study, we want to help the
community to better understand the fundamental causes of compression
performance in the field of waveform-based whole-building energy datasets. We
provide a definition of measurement calibration and its effects on the storage
requirements based on signal entropy. Published datasets are self-contained and
final, which allows us to prioritize the compression ratio and achievable space
saving over other common compression metrics (CPU load, throughput, or latency).
We define the achievable space saving and compression ratio as the only
criterion when dealing with large (offline) datasets.

The rest of this paper is structured as follows: We discuss related work in
Section~\ref{sec:related-work}. We describe the evaluated datasets in
Section~\ref{sec:evaluated-datasets}, which are then used in the experiments in
Sections~\ref{sec:entropy-analysis},~\ref{sec:data-representation},
and~\ref{sec:chunk-size-impact}. Finally, we present results in
Section~\ref{sec:results}, before concluding in Section~\ref{sec:conclusions}.

%% file: 02-related-work.tex

\section{Related Work}
\label{sec:related-work}

NILM and related fields distinguish between \emph{low} and \emph{high sampling
rates} to capture voltage and current measurements. Low sampling rates (or
low-frequency) are typically \SI{1}{\Hz} or slower. High sampling rates (or
high-frequency) are typically above \SI{500}{\Hz} (or at least the
Nyquist–Shannon sampling theorem \cite{Shannon1949}). Recording  multiple
channels with high sampling rates requires oscilloscopes or specialized data
acquisition systems as presented in \cite{Kriechbaumer2017-MEDAL, Haq2017-CLEAR,
Meziane2016}.

Low-frequency energy data can benefit greatly from compression when applied to
smart meter data, as multiple recent works have shown
\cite{Ringwelski2012,Unterweger2015a,Unterweger2015b,Eichinger2015}. Electricity
smart meters can be a source of high data volume with measurement intervals of
\SI{1}{\second}, \SI{60}{\second}, \SI{15}{\minute}, or higher. Possible
transmission and storage savings due to lossless compression have been evaluated
in \cite{Unterweger2015b}. While the achievable compression ratio increased with
smaller sampling intervals, the benefits of compression vanish quickly above
\SI{15}{\minute} intervals. Various encodings (ASCII- and binary-based) have
been evaluated for such low-frequency measurements, and in most cases, a binary
encoding greatly outperforms an ASCII-based encoding. The need for smart data
compression was discussed in \cite{Nabeel2013}, which further motivates in-depth
research in this area. The main focus of the authors was smart meter data with
low temporal resolution from 10,000 meters or more. Various compression
techniques were presented and a fast-streaming differential compression
algorithm was evaluated: removing steady-state power measurements ($t_{i+1} -
t_{i} = 0$) can save on average 62\% of required storage space.

High-frequency energy data offers a significantly larger potential for lossless
compression, due to the inherent repeating waveform signal. \citet{Tariq2015}
utilized general-purpose compressors, such as LZMA and bzip2, and achieved good
compression ratios on some datasets. Applying differential compression and
omitting timestamps can yield size reductions of up to 98\% on smart grid data,
however, these results are not comparable as there is no generalized uniform
data source. The presented results use a single data channel and an ASCII-based
data representation as a baseline for their comparison, which contains an inherent
encoding overhead. The SURF file format \cite{Pereira2014} was designed to
store NILM datasets and provide an API to create and modify such files. The
internal structure is based on wave-audio and augments it with new types of
metadata chunks. To the best of our knowledge, the SURF file format didn't gain
any traction due to its lack of support in common scientific computing
frameworks. The recently published EMD-DF file format \cite{Pereira2017-EMD-DF},
by the same authors, relies on the same wave-audio encoding, while extending it
with more metadata and annotations. Neither SURF nor EMD-DF provides any built-in
support for compression. The power grid community defined the PQDIF \cite{PQDIF}
(for power quality and quantity monitoring) and COMTRADE \cite{COMTRADE} (for
transient data in power systems) file formats.  Both specifications outline a
structured view of numerical data in the context of energy measurements. Raw
measurements are augmented with precomputed evaluations (statistical metrics),
which can cause a significant overhead in required storage space. While PQDIF
supports a simple LZ compression, COMTRADE does not offer such capabilities. To
the best of our knowledge, these file formats never gained traction outside the
power grid operations community.

Lossy compression can achieve multiple magnitudes higher compression ratios than
lossless, with minimal loss of accuracy for certain use cases
\cite{Eichinger2015}. Using piecewise polynomial regression, the authors
achieved good compression ratios on three existing smart grid scenarios. The
compressed parametrical representation was stored in a relational database
system. However, this approach only applies if the use case and expected data
transformation is known before applying a lossy data reduction. A 2-dimensional
representation for power quality data was proposed in \cite{Gerek2004} and
\cite{Qing2011}, which then could be used to employ compression approaches from
image processing and other related fields. While both approaches can be
categorized as lossy compression due to their numerical approximation using
wavelets or  trigonometric functions, they require a specialized encoder and
decoder which is not readily available in scientific computing frameworks.

The NilmDB project \cite{Paris2014} provides a generalized user interface to
access, query, and analyze large time-series datasets in the context of power
quality diagnostics and NILM. A distributed architecture and a custom storage
format were employed to work efficiently with ``big data''. The underlying data
persistence is organized hierarchically in the filesystem and utilizes
tree-based structures to reduce storage overhead. This internal data
representation is capable of handling multiple streams and non-uniform data
rates but lacks support for data compression or more efficient coding schemes.
NILMTK \cite{NILMTK}, an open-source NILM toolkit, provides an evaluation
workbench for power disaggregation and uses the HDF5 \cite{Folk2011} file format
with a custom metadata structure. Most available public datasets require a
specialized converter to import them into a NILMTK-usable file format. While the
documentation states that a zlib data compression is applied, some
converters currently use bzip2 or Blosc \cite{Blosc}.

%% file: 03-evaluated-datasets.tex

\section{Evaluated Datasets}
\label{sec:evaluated-datasets}

While there is a vast pool of smart meter datasets\footnotemark[1], i.e., low sampling rates of
measurements every \SI{1}{\second}, \SI{15}{\min}, or \SI{1}{\hour}, a majority
of the underlying information is already lost (signal waveform). The raw signals
are aggregated into single root-mean-squared voltage and current readings,
frequency spectrums, or other metrics accumulated over the last measurement
interval. This can be already classified as a type of lossy compression. For
some use cases, this data source is sufficient to work with, while other fields
require high sampling rates to extract more information from the signals.

All following experiments and evaluations were performed on publicly
accessible datasets: \emph{The Reference Energy Disaggregation Data Set} (REDD
\cite{Kolter2011-REDD}), \emph{Building-Level fUlly-labeled dataset for
Electricity Disaggregation} (BLUED \cite{Anderson2012-BLUED}), \emph{UK
Domestic Appliance-Level Electricity dataset} (UK-DALE
\cite{Kelly2015-UK-DALE}), and the \emph{Building-Level Office eNvironment
Dataset} (BLOND \cite{Kriechbaumer2018-BLOND}). We will refer to these datasets
by their established acronyms: REDD, BLUED, UK-DALE, and BLOND. Based on the
energy dataset survey provided by the
NILM-Wiki\footnotemark[1], these are all
datasets of long-term continuous measurements with voltage and current waveforms
from selected buildings or households. The data acquisition systems and data
types are comparable to warrant their use in this context.
(Table~\ref{tab:datasets}).

\footnotetext[1]{\url{http://wiki.nilm.eu/datasets.html}}

Measurement systems and their analog-to-digital converters (ADC) always output a
unit-less integer number, either between $[0, 2^{bits})$ for unipolar ADCs or
$[-2^{bits-1}, 2^{bits-1})$ for bipolar ADCs. During setup and calibration, a
common factor is determined to convert raw values into a voltage or current
reading. Some datasets publish raw values and the corresponding calibration
factors, while others publish directly Volt- and Ampere-based readings as float
values. Datasets only available as floating-point values are converted back into
their original integer representation without loss of precision by reversing the
calibration step from the analog-to-digital converter for each channel:%
\begin{align*}
  measurement_i &= ADC_i \cdot calibration_{channel} \\
  [Volts] &= [steps] \cdot [\nicefrac{Volt}{steps}] \\
  [Ampere] &= [steps] \cdot [\nicefrac{Ampere}{step}]
\end{align*}

Each of the mentioned datasets was published in a different (compressed) file
format and encoding scheme. To allow for comparisons between these datasets, we
decompressed, normalized, and re-encoded all data before analyzing them (raw
binary encoding).

From REDD, we used the entire available \emph{High Frequency Raw Data}:
\emph{house\_3} and \emph{house\_5}, each with 3 channels:
\emph{current\_1}, \emph{current\_2}, and \emph{voltage}. The custom file format
encodes a single channel per file. In total, \SI{1.4}{\gibi\byte} of raw data
from 126 files were used.

From BLUED, we used all available waveform data (1 location, 16 sub-datasets)
and 3 channels: \emph{current\_a}, \emph{current\_b}, \emph{voltage}. The
CSV-like text files contain voltage and two current channels and a dedicated
measurement timestamp. In total, \SI{41.1}{\gibi\byte} of raw data from 6430
files were used.

From UK-DALE, we selected \emph{house\_1} from the most recent release
(\emph{UK-DALE-2017-16kHz}, the longest continuous recording). The compressed
FLAC files contain 2 channels: \emph{current} and \emph{voltage}. In total,
\SI{6259.1}{\gibi\byte} of raw data from 19491 files were used.

From BLOND, we selected the aggregated mains data of both sub-datasets: BLOND-50
and BLOND-250. The HDF5 files with gzip compression contain 6 channels:
\emph{current\{1-3\}} and \emph{voltage\{1-3\}}. In total,
\SI{10246.7}{\gibi\byte} of raw data from 61125 files of BLOND-50, and
\SI{11899.0}{\gibi\byte} of raw data from 35490 files of BLOND-250 were used.

\begin{table}[!tbp]
  \centering
  \caption{Overview of evaluated datasets: long-term continuous measurements containing raw voltage and current waveforms.}
  \label{tab:datasets}
  \begin{tabular}{lccrc}
    \toprule
    \textbf{Dataset} & \multicolumn{1}{>{\centering\arraybackslash}m{39pt}}{\textbf{Current Channels}} & \multicolumn{1}{>{\centering\arraybackslash}m{39pt}}{\textbf{Voltage Channels}} & \multicolumn{1}{>{\centering\arraybackslash}m{37pt}}{\textbf{Sampling Rate}} & \multicolumn{1}{>{\centering\arraybackslash}m{32pt}}{\textbf{Values}} \\ \midrule
    REDD       & 2 & 1 & \SI{15}{\kilo\Hz}  & 24-bit \\
    BLUED      & 2 & 1 & \SI{12}{\kilo\Hz}  & 16-bit \\
    UK-DALE    & 1 & 1 & \SI{16}{\kilo\Hz}  & 24-bit \\
    BLOND-50   & 3 & 3 & \SI{50}{\kilo\Hz}  & 16-bit \\
    BLOND-250  & 3 & 3 & \SI{250}{\kilo\Hz} & 16-bit \\
    \bottomrule
  \end{tabular}
\end{table}

The data acquisition systems (DAQ) of all datasets produce a linear pulse-code
modulated (LPCM) stream. The analog signals are sampled in uniform intervals
and converted to digital values (Figure~\ref{fig:pcm-stream}). The quantization
levels are distributed linearly in a fixed measurement range which requires a
signal conditioning step in the DAQ system. ADCs
typically cannot directly measure mains voltage and require a step-down
converter or measurement probe. Mains current signals need to be converted into
a proportional voltage.

%% file: 04-entropy-analysis.tex

\section{Entropy Analysis}
\label{sec:entropy-analysis}

DAQ units provide a way to collect digital values from
analog systems. As such, the quality of the data depends strongly on the correct
calibration and selection of measurement equipment. Mains electricity signals
are typically not compatible with modern digital systems, requiring an indirect
measurement through step-down transformers or other metrics. Mains voltage can
vary by up to $\pm10\%$ during normal operation of the grid
\cite{CENELEC:HD472S1,ANSI:C84.1}, making it necessary to design the measurement
range with a safety margin. The expected signal, plus any margin for spikes,
should be equally distributed on the available ADC resolution range. Leaving
large areas of the available value range unused can be prevented by carefully
selecting input characteristics and signal conditioning (step-down calibration).
A rule of thumb for range calibration is that the expected signal should occupy
80-90\%, leaving enough bandwidth for unexpected measurements. Input signals
larger than the measurement range get recorded as the minimum/maximum value.
Grossly exceeding the rated input signal level could damage the ADC, unless a
dedicated signal conditioning and protection is employed.

\begin{figure}[!htbp]
  \centering
  \includegraphics[width=0.35\textwidth]{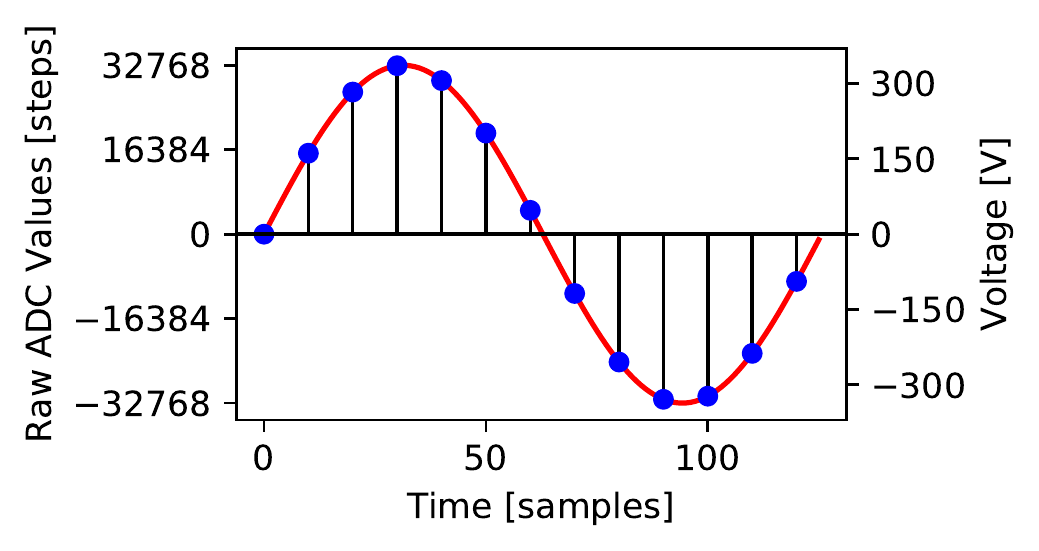}
  \caption{Linear pulse-code modulation stream of a sinusoidal waveform sampled with a 16-bit ADC. The waveform corresponds to a 230\,V mains voltage signal.}
  \label{fig:pcm-stream}
\end{figure}

We extracted the probability mass function (PMF) of all evaluated datasets for
the full bit-range (16- or 24-bit). The value histogram is a structure mapping
each possible measurement value (integer) to the number of times this value was
recorded. Ideally, the region between the lowest and highest value contains a
continuous value range without gaps. However, the quantization level (step size)
could cause a mismatch and results in skipped values.  We then normalize this
histogram to obtain the PMF and compute the signal entropy per channel, which
gives an estimation of the actual information contained in the raw data and
provides a lower bound for the achievable compression ratio based on the
Kolmogorov complexity. 
\begin{align*}
  X &= \left\{ -2^{bits-1}, ..., 0, ..., 2^{bits-1}-1 \right\} \\
  hist &= histogram(dataset, X) \\
  f_X &= \frac{hist}{\sum_{x \in X} hist[x]} \\
  \forall x \in X &\text{ where } f_X(x) = 0: f_X(x) = 1 \\
  H(x) &= - \sum_{x \in X} f_X(x) \cdot log_2\left( f_X(x) \right)
\end{align*}

Each dataset is split into multiple files, making it necessary to merge all
histograms into a total result at the end of the computing run. Since all
histograms can be combined with a simple summation, the process can be
parallelized and computed without any particular order. Computing and merging
all histograms is, therefore, best accomplished in a distributed compute cluster
with multiple nodes or similar environments.

%% file: 05-data-representation.tex

\section{Data Representation}
\label{sec:data-representation}

Choosing a suitable file format for potentially large datasets involves multiple
tradeoffs and decisions, including supported platforms, scientific computing
frameworks, metadata, error correction, compression, and chunking. The available
choices for data representation can range from CSV data (ASCII-parsable) to
binary file formats and custom encoding schemes. From the energy dataset survey
and the evaluated datasets, it can be noted, that every dataset uses a different
file format, encoding scheme, and optionally compression.

Publishing and distributing large datasets requires storage systems capable of
providing long-term archives of scientific measurement data. Lossless
compression helps to minimize storage costs and distribution efforts. At the
same time, other researchers accessing the data benefit from smaller files and
shorter access times to download the data.

Electricity signals (current and voltage) contain a repetitive waveform with
some form of distortion depending on the load. In an ideal power grid, the
voltage would follow a perfect sinusoidal waveform without any offset or error.
This would allow us to accurately predict the next voltage measurement. However,
constant fluctuations in the supply and demand cause the signals to deviate. The
fact that each signal is primarily continuous (without sudden jumps) can be
beneficial to compression algorithms.

A delta encoding scheme only stores the numerical difference of
neighboring elements in a time-series measurement vector. This can be useful for
slow-changing signals because the difference of a signal might require less
bytes to encode than the absolute value:%
\begin{align*}
  \forall i \in \left\{ 1 \dots n \right\}: d_i &= v_i - v_{i-1} \\
  d_0 &= v_0
\end{align*}

We compare the original data representation (format, compression, encoding) of
each dataset, reformat them into various file formats, and evaluate their
storage saving based on a comprehensive list of lossless compression algorithms.
This involves encoding raw data in a more suitable representation to compare
their compressed size: $CS = compressed\_size / original\_size * 100\%$, and the
resulting space saving: $SS = 100\% - CS$. We define the main goal of reducing
the overall required storage space for each dataset, and deliberately do not
consider compression or decompression speed. The performance characteristics
(throughput and speed) are well known for individual compression techniques
\cite{Arnold1997} and are of minor importance in the case of large static
datasets which require only a single compression step before distribution.
Performance metrics are important when dealing with repeated compression of raw
data, which is not the case for static energy datasets. Repeated decompression
is however relevant because researchers might want to read and parse the files
over and over again while analyzing them (if in-memory processing is not
feasible). As noted in \cite{Arnold1997}, decompression speed and throughput is
typically not a performance bottleneck in data analytics tasks.

Building a novel data compression scheme for energy data is counter-productive,
since most scientific computing frameworks lack support and the idea suffers
from the "not invented here" and "yet another standard" problematic, both common
anti-patterns in the field of engineering when developing new solutions, despite
existing suitable approaches \cite{Pereira2014, Kolter2011-REDD, Eichinger2015}.
Therefore, a key requirement is that each file format must be supported in
common scientific computing systems to read (and possibly write) data files.

We selected four format types: raw binary, HDF5 (data model and file format for
storing and managing data), Zarr (chunked, compressed, N-dimensional arrays),
and audio-based PCM containers.

Raw binary formats provide a baseline for comparison. All samples are encoded as
integer values (16-bit or 24-bit) and are compressed with a general-purpose
compressor: zlib/gzip, LZMA, bzip2, and zstd, all with various parameter values.
The input for each compressor is either raw-integer or variable-length encoded
data (LEB128S \cite{DWARF3.0}), which is serialized either row- or column-based
from all channels (interweaving). The LEB128S encoding is additionally evaluated
with delta encoding of the input.

The Hierarchical Data Format 5 (HDF5) \cite{Folk2011} provides structured
metadata and data storage, data transformations, and libraries for most
scientific computing frameworks. All data is organized in natively-typed arrays
(multi-dimensional matrices) with various filters for data compression,
checksumming, and other reversible transformations before storing the data to a
file. The API transparently reverses these transformations and compression
filters while reading data. HDF5 is popular in the scientific community and used
for various big-data-type applications
\cite{Blanas2014,Gosink2006,Dougherty2009,Sehrish2017}. The public registry for
HDF5 filters\footnote{\url{https://support.hdfgroup.org/services/filters.html}}
currently lists 21 data transformations, most of them compression-related. Each
HDF5 file is evaluated with and without the shuffle filter, zlib/gzip, lzf,
MAFISC \cite{Huebbe2013} with LZMA, szip \cite{szip}, Bitshuffle
\cite{Masui2015} with LZ4, zstd, and the full Blosc \cite{Blosc}
compression suite, again all with various parameter values.

Zarr \cite{Zarr} organizes all data in a filesystem-like structure, which can be
archived as a single zip-archive file or as tree-structure in the filesystem. Each
channel is stored as a separate array (data stream) with optional chunk-based
compression via zlib/gzip, LZMA, bzip2, or Blosc (with shuffle, Bitshuffle, or
no-shuffle filter), again all with various parameter values. Each Zarr file is
additionally evaluated with a delta filter to reduce the value range.

Audio-based formats use LPCM-type data encoding (\texttt{PCM16} or
\texttt{PCM24}) with a fixed precision and sampling rate. All channels are
encoded into a single container using lossless compression formats: FLAC
\cite{FLAC}, ALAC \cite{ALAC}, and WavPack \cite{WavPack}. These formats do not
provide tune-able parameters.

Calibration factors, timestamps, and labels can augment the raw data in a single
file while providing a unified API for accessing data and metadata. Raw binary
formats lack this type of integrated support and require additional tooling and
encoding schemes for metadata. Audio-based formats require a container format to
store metadata, typically designed for the needs of the music and entertainment
industry. Out of these formats, only HDF5 and Zarr provide support for encoding
and storing arbitrary metadata objects (complex types or matrices) together with
measurement data.

Most audio-based formats support at most 8 signal channels, while
general-purpose formats such as HDF5 and Zarr have no restrictions on the total
number of channels per file. The sampling rate can also be a limiting factor: FLAC
supports at most \SI{655.35}{\kilo\Hz} and ALAC only \SI{384}{\kilo\Hz}. ADC
resolution (bit depth) is mostly bound by existing technological limitations and
will not exceed 32-bit in the foreseeable future. While these constraints are
within the requirements for all datasets under evaluation, they need to be
considered for future dataset collection and the design of measurement systems.

In total, we encoded the evaluated datasets with \numberOfDataRepresentations{}
different data representation formats: 54 raw, 264 HDF5-based, 44 Zarr-based,
and 3 audio-based and gathered their per-file compression size as a benchmark.
The complete list, including all parameters and compression options, is
available in the online appendix\footnote{The online appendix is available
through the program chair (double-blind review).}. The full analysis was
performed in a distributed computing environment and consumed approx.
$1,176,000$ CPU-core-hours (dual Intel Xeon E5-2630v3 machines with
\SI{128}{\gibi\byte} RAM and \SI{10}{\gibi\bit} Ethernet interfaces).

%% file: 06-chunk-size-impact.tex

\section{Chunk Size Impact}
\label{sec:chunk-size-impact}

Each dataset is provided in equally-sized files, typically based on measurement
duration. Working with a single large file can be cumbersome due to main memory
restriction or available local storage space. Assuming a typical desktop
computer, with \SI{8}{\gibi\byte} of main memory, is used for processing, a
single file from a dataset must be fully loaded into memory before any
computation can be done. Depending on the analysis and algorithms, multiple
copies might be required for intermediary results and temporary copies. This
means the main memory size is an upper bound for the maximum feasible chunk
size.

Some file formats and data types support internal chunking or streamed data
access, in which data can be read into memory sequentially or random-access. In
such environments other factors will limit the usable chunk size, such as file
system capabilities, network-attached storage, or other operating system
limitations.

The evaluated datasets are distributed with the following chunk sizes of raw
data: REDD: \SI{11.4}{\mebi\byte} or \SI{4}{\minute}, BLUED:
\SI{6.6}{\mebi\byte} or \SI{1.65}{\minute}, UK-DALE: \SI{329.2}{\mebi\byte} or
\SI{60}{\minute}, BLOND-50: \SI{171.7}{\mebi\byte} or \SI{5}{\minute},
BLOND-250: \SI{343.3}{\mebi\byte} or \SI{2}{\minute}. Measurement duration and
file size are not strictly linked, causing a slight variation in file sizes
across the entire measurement period of each dataset. Observed real-world time
does not affect any of the compression algorithms under test and is therefore
omitted. The sampling rate and channel count directly affects the data rate
(bytes per time unit) and explains the non-uniform chunk sizes mentioned for
each dataset.

We compare the best-performing data representation formats of each dataset from
the previous experiment, benchmark them with different chunk sizes, and estimate
their effect on the overall compression ratio. For this evaluation, we define
the compression ratio as $CR = original\_size / compressed\_size$. The chunk
sizes range from 1, 2, 4, 8, 16, 32, 64, \SI{128}{\mebi\byte}, and then
continue in steps of \SI{128}{\mebi\byte} up to \SI{3072}{\mebi\byte}. To
reduce the required computational effort, we greedily consume data from the
first available dataset file, until the predefined chunk limit is fulfilled. The
chunk size is determined using the number of samples (across all channels) and
their integer byte count (2 or 3 bytes); only full samples for all channels are
included in a chunk.

%% file: 07-results.tex

\section{Results}
\label{sec:results}

\subsection{Entropy Analysis}
\label{sec:results-entropy-analysis}

\begin{figure*}[p]
  \centering
  \begin{subfigure}{0.475\textwidth}
    \centering
    \includegraphics[width=\textwidth]{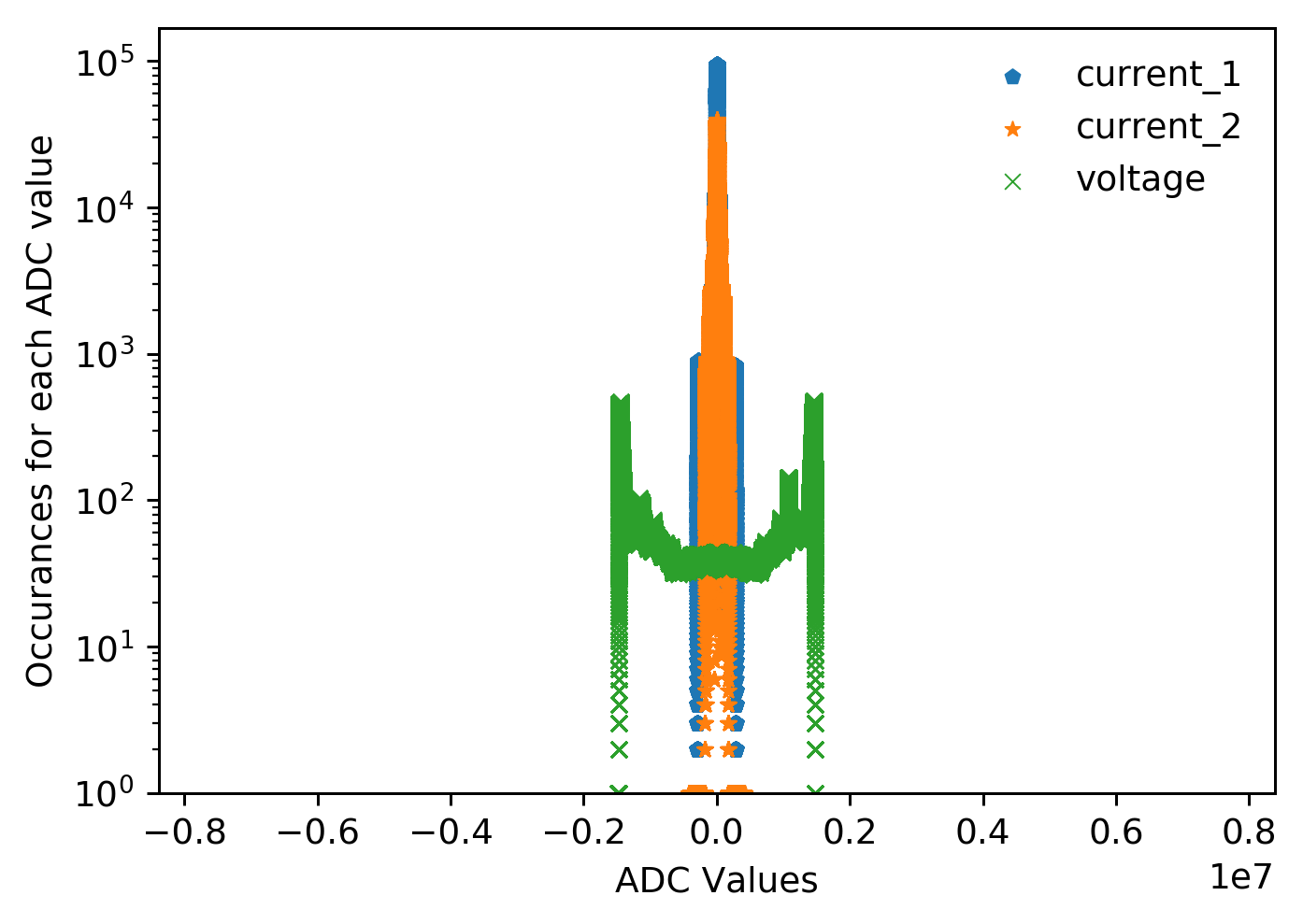}
    \subcaption{REDD (24-bit)}
    \label{fig:histogram-adc-REDD}
  \end{subfigure}
  \begin{subfigure}{0.475\textwidth}
    \centering
    \includegraphics[width=\textwidth]{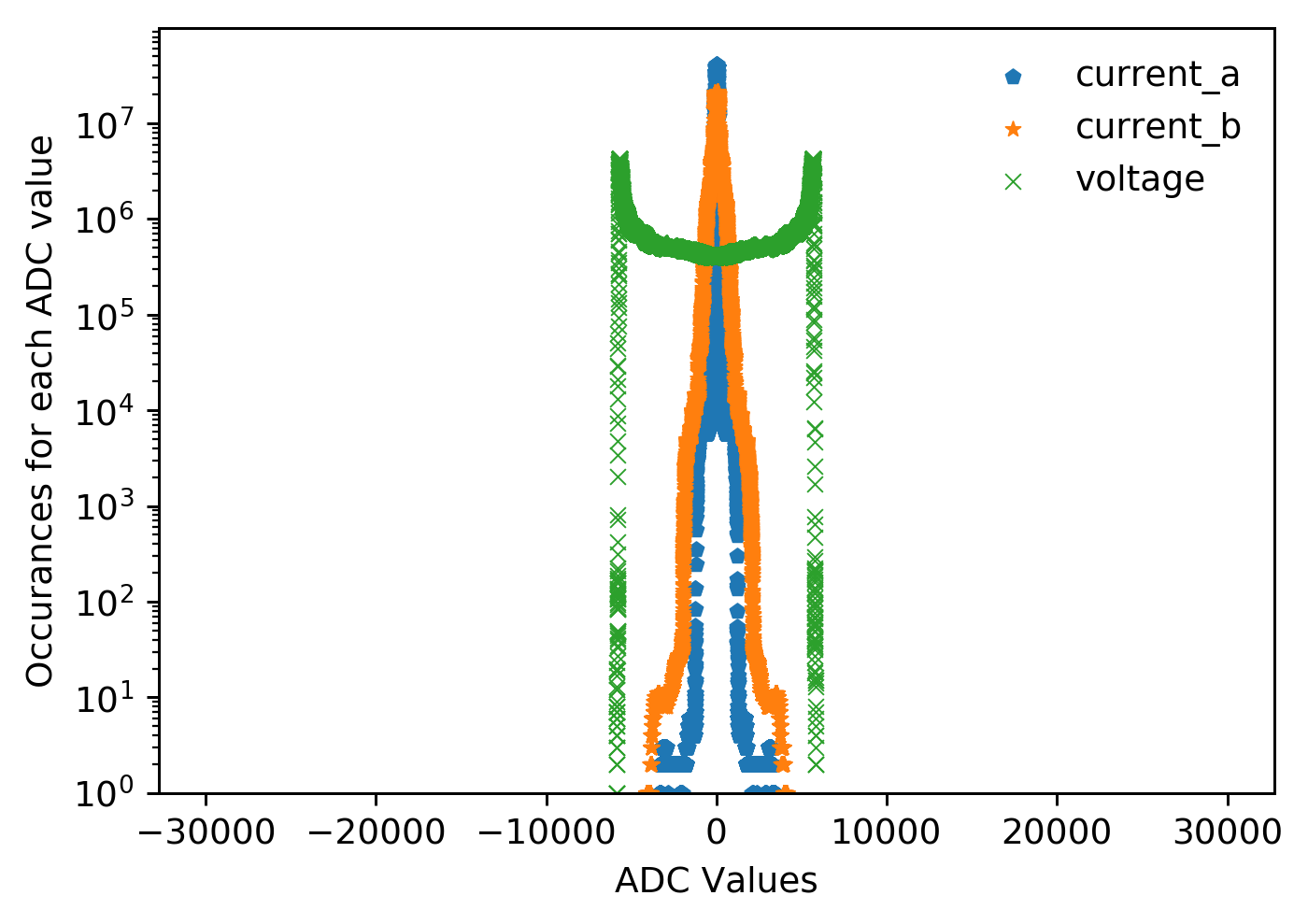}
    \subcaption{BLUED (16-bit)}
    \label{fig:histogram-adc-BLUED}
  \end{subfigure}
  \begin{subfigure}{0.475\textwidth}
    \centering
    \includegraphics[width=\textwidth]{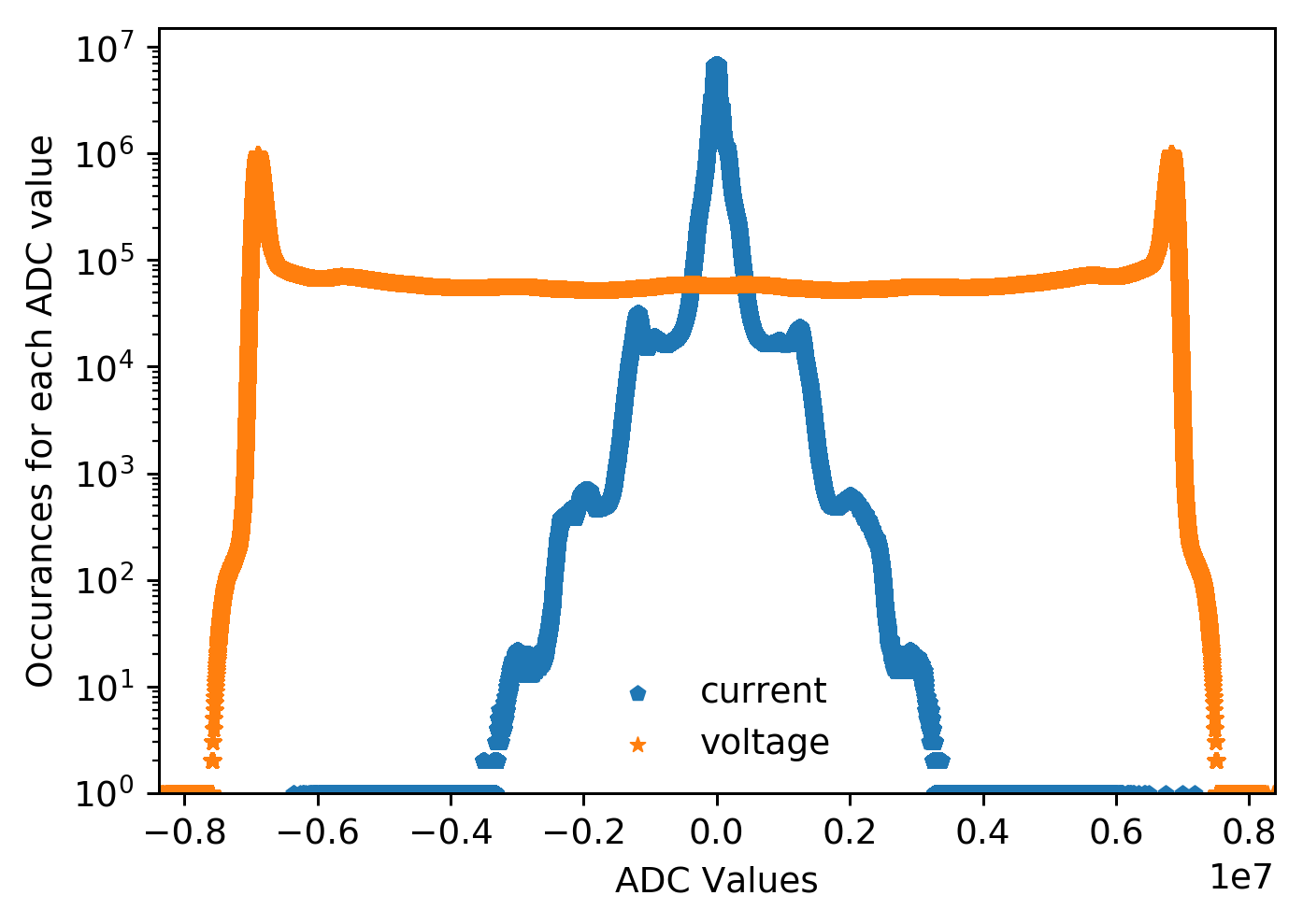}
    \subcaption{UK-DALE (24-bit)}
    \label{fig:histogram-adc-UKDALE}
  \end{subfigure}
  \begin{subfigure}{0.475\textwidth}
    \centering
    \includegraphics[width=\textwidth]{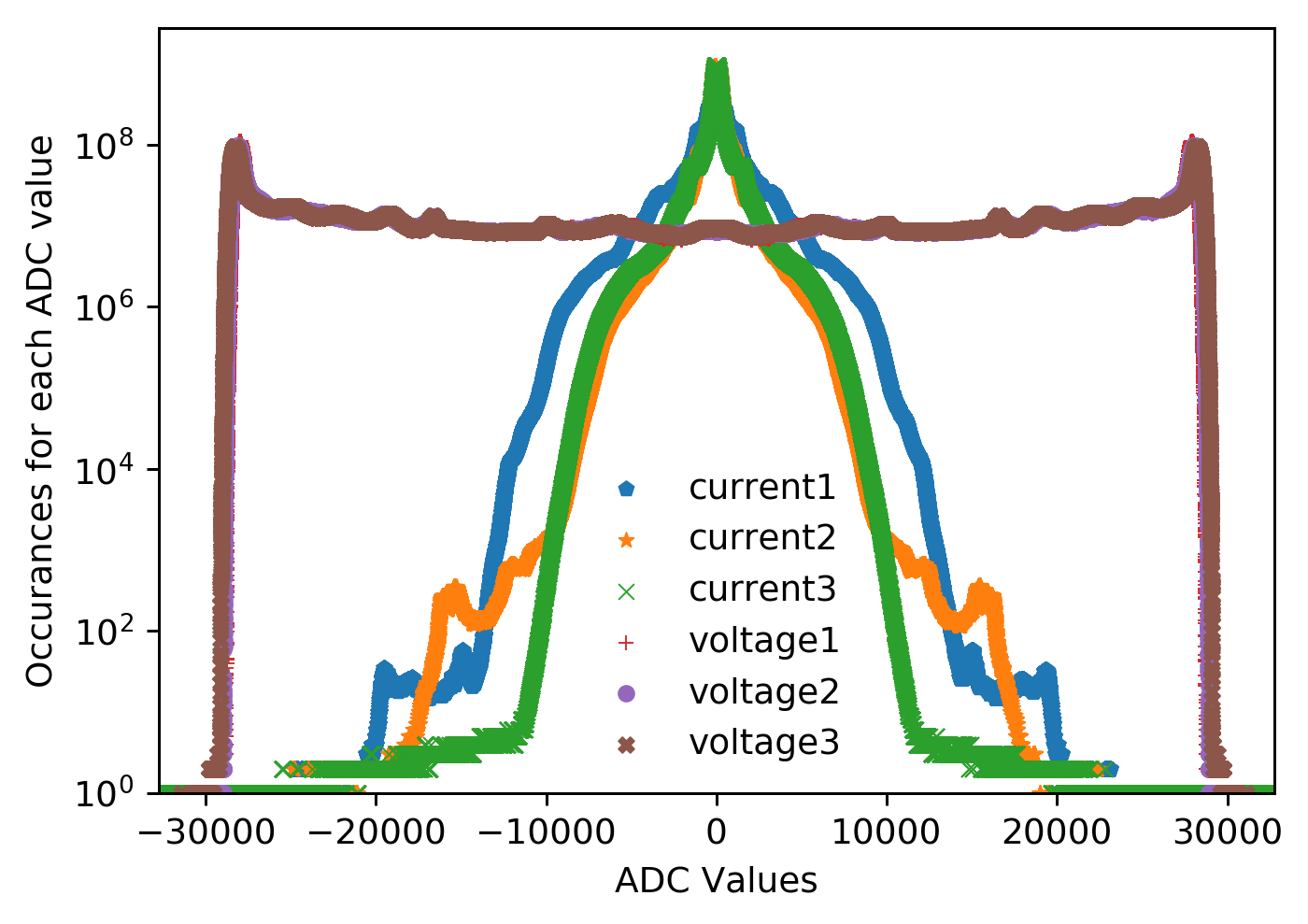}
    \subcaption{BLOND-50 (16-bit)}
    \label{fig:histogram-adc-BLOND50}
  \end{subfigure}
  \begin{subfigure}{0.475\textwidth}
    \centering
    \includegraphics[width=\textwidth]{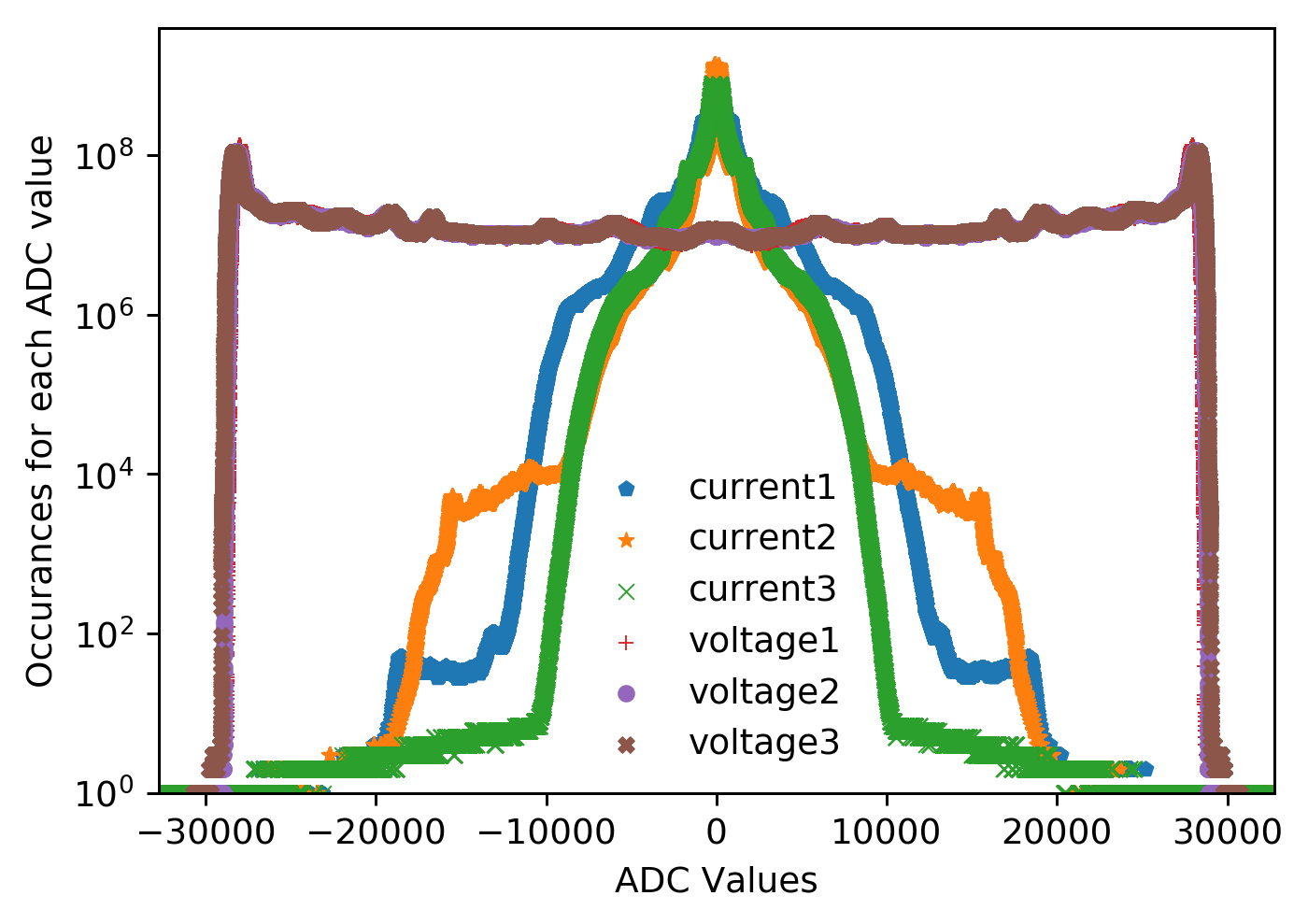}
    \subcaption{BLOND-250 (16-bit)}
    \label{fig:histogram-adc-BLOND250}
  \end{subfigure}
  \caption[Semi-logarithmic histogram of ADC values for each dataset and channel.]{
  Semi-logarithmic histogram of ADC values for each dataset and channel.
  Current signals show distinct steps, corresponding to prolonged usage at certain power levels.
  For visualization reasons, the scatter plot was smoothed and the full histogram is available in the online appendix\footnotemark[3].
  }
  \label{fig:histogram-adc}
\end{figure*}

Entropy is based on the probability for a given measurement (signal value). The
histogram of an entire measurement channel shows the number of times a single
measurement value was seen in the dataset (Figure~\ref{fig:histogram-adc}). The
plots show the raw measurement bandwidth in ADC value on the x-axis and a
logarithmic y-axis for the number of occurrences of each value. The raw ADC
values are bipolar and centered on 0: $-32768 \ldots 32767$ for BLUED, BLOND-50,
and BLOND-250; $-8388608 \ldots 8388607$ for REDD and UK-DALE.

The voltage histogram shows a distinctive sinusoidal distribution (peaks at
minimum and maximum values). The current histogram would show a similar
distribution if the power draw is constant (pure-linear or resistive loads),
however, multiple levels of current values can be observed, indicating high
activity and fluctuations. REDD and BLUED (Figures~\ref{fig:histogram-adc-REDD}
and~\ref{fig:histogram-adc-BLUED}) show a center-biased distribution, indicating
a sub-optimal calibration performance and unused measurement bandwidth. UK-DALE,
BLOND-50, and BLOND-250 (Figures~\ref{fig:histogram-adc-UKDALE},
\ref{fig:histogram-adc-BLOND50}, \ref{fig:histogram-adc-BLOND250}) show a wide
range of highly used values, with the voltage channels utilizing around 90\% of
the available bandwidth.

\begin{figure*}[t]
  \centering
  \includegraphics[width=\textwidth]{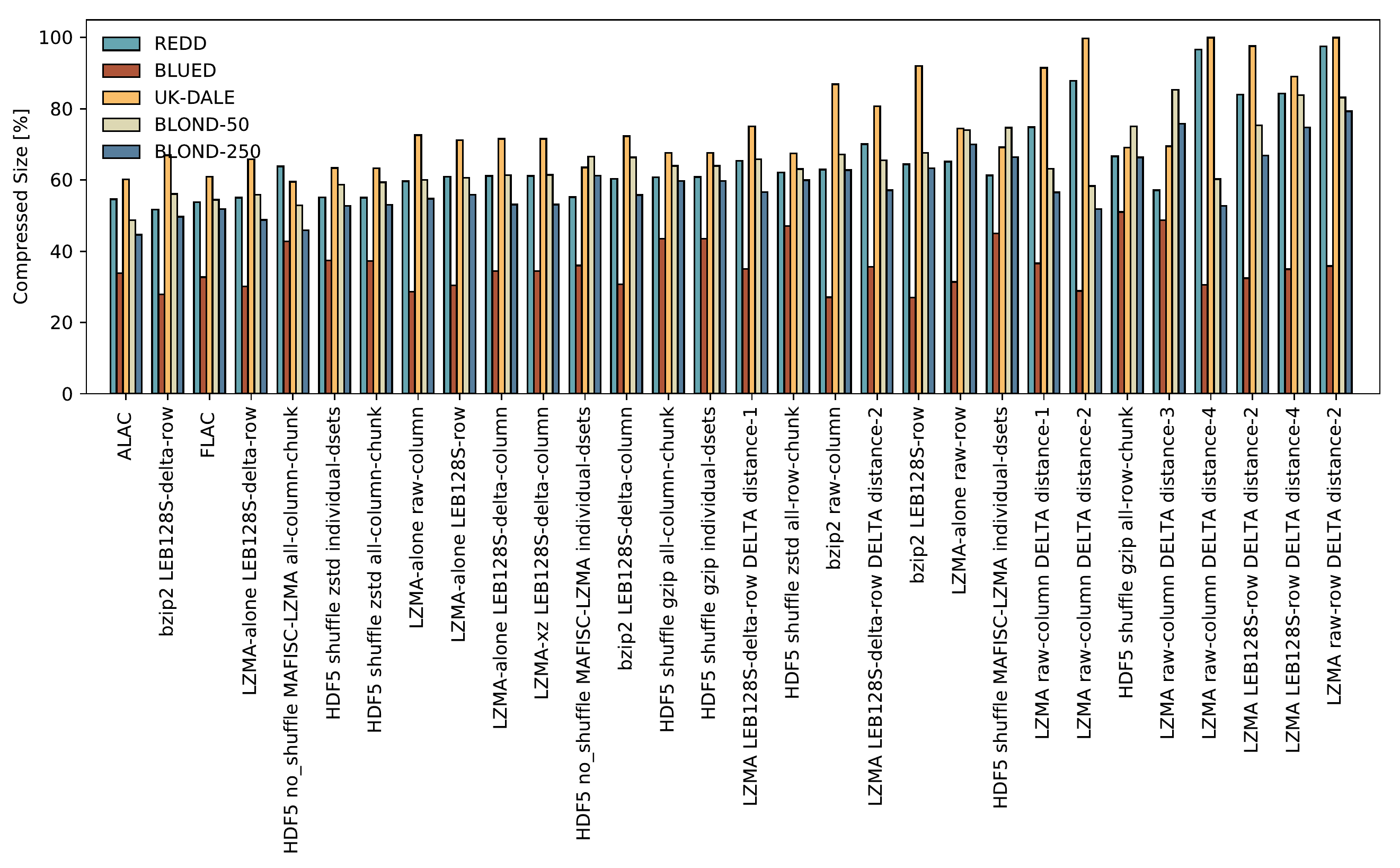}
  \caption{Compression performance for the top-30 data representation formats and their transformation filters. Each data representation format was applied on a per-file basis to every dataset.}
  \label{fig:file-format-eval}
\end{figure*}

REDD and BLUED use only a small percentage of the available range, indicating a
low entropy based on the used data type. UK-DALE utilizes a reasonable slice,
while BLOND covers almost the entire possible range
(Table~\ref{tab:entropy-used-values}). Assuming a well-calibrated data
acquisition system, the expected percentage should reflect the expected
measurement values. Low range usage (REDD, BLUED) leads to lost precision which
would have been freely available with the given hardware, whereas high usage
(UK-DALE, BLOND) means almost all available measurement precision is reflected
in the raw data. Some datasets utilize 100\% of the available measurement range,
while REDD only uses 5\%. A high range utilization does not result in a equally
high usage, as the histogram can contain gaps (ADC values with 0 occurrences in
the datasets).

\begin{table}
  \centering
  \caption{Entropy analysis of whole-building energy datasets with high sampling
  rates. The amount of unique measurement values for each channel is extracted, which corresponds
  to a usage percentage over the available measurement resolution. The lowest and
  highest observed value is used to give determine the observed range.}
  \label{tab:entropy-used-values}
  \begin{tabular}{ccrrrr}
  \toprule
  \textbf{Dataset} & \textbf{Channel} & \textbf{Values} & \textbf{Usage} & \textbf{Range} & \textbf{H(x)}\\
  \midrule
  \multirow[c]{3}{16mm}{\centering REDD \\ (24-bit)}      & current\_1 &    87713 &    1\% &     4\% &  14.3 \\
                                                          & current\_2 &    85989 &    1\% &     5\% &  14.9 \\
                                                          &    voltage &  2925155 &   17\% &    18\% &  21.1 \\ \midrule
  \multirow[c]{3}{16mm}{\centering BLUED \\ (16-bit)}     & current\_a &     5855 &    9\% &    10\% &   7.8 \\
                                                          & current\_b &     7684 &   12\% &    13\% &   9.7 \\
                                                          &    voltage &    11302 &   17\% &    18\% &  13.2 \\ \midrule
  \multirow[c]{2}{16mm}{\centering UK-DALE \\ (24-bit)}   &    current &  6981612 &   42\% &    81\% &  19.0 \\
                                                          &    voltage & 15135594 &   90\% &   100\% &  23.2 \\ \midrule
  \multirow[c]{6}{16mm}{\centering BLOND-50 \\ (16-bit)}  &   current1 &    51122 &   78\% &   100\% &  12.6 \\
                                                          &   current2 &    49355 &   75\% &   100\% &  11.2 \\
                                                          &   current3 &    48658 &   74\% &   100\% &  11.3 \\
                                                          &   voltage1 &    58396 &   89\% &    92\% &  15.3 \\
                                                          &   voltage2 &    57975 &   88\% &    91\% &  15.4 \\
                                                          &   voltage3 &    59596 &   91\% &    95\% &  15.4 \\ \midrule
  \multirow[c]{6}{16mm}{\centering BLOND-250 \\ (16-bit)} &   current1 &    52721 &   80\% &   100\% &  12.4 \\
                                                          &   current2 &    51802 &   79\% &   100\% &  10.8 \\
                                                          &   current3 &    50989 &   78\% &   100\% &  11.6 \\
                                                          &   voltage1 &    58488 &   89\% &    91\% &  15.3 \\
                                                          &   voltage2 &    57912 &   88\% &    92\% &  15.4 \\
                                                          &   voltage3 &    59742 &   91\% &    94\% &  15.4 \\
  \bottomrule
  \end{tabular}
\end{table}

\subsection{Data Representation}
\label{sec:results-data-representation}

The evaluation compares the compressed size (CS, final file size after
compression and file format encapsulation in percent of uncompressed size) of
\numberOfDataRepresentations{} data representation formats. For brevity reasons,
only the 30 best-performing formats are shown in
Figure~\ref{fig:file-format-eval}. Each of the \numberOfDataRepresentations{}
data representation was tested on all datasets and the full evaluation is
available in the online appendix\footnotemark[3]. The following
evaluation and benchmark uses the raw data from each dataset as described in
Section~\ref{sec:evaluated-datasets}. In total, raw data with
\SI{27.8}{\tebi\byte} was re-encoded \numberOfDataRepresentations{} times.

\footnotetext[3]{The online appendix is available through the program chair (double-blind review).}

HDF5 and Zarr are general-purpose file formats for numerical data with a broad
support in scientific computing frameworks. As such, they only support 16-bit
and 32-bit integer values, which causes a 1-byte overhead for REDD and UK-DALE.
The baseline used for comparison is a raw concatenated byte string with
dataset-native data types (16-bit and 24-bit). This allows us to obtain
comparable evaluation results, while other published benchmarks compared
ASCII-like encodings against binary representations, skewing the results
significantly.

Overall, it can be noted that all three audio-based formats performed well,
given their inherent targeted nature of compressing waveforms with high temporal
resolution. ALAC and FLAC achieved the highest overall CS across all datasets,
followed by HDF5+MAFISC and HDF5+zstd, which can overcome the 1-byte overhead.
Although the general-purpose compressors and their individual data
representation formats were intended to serve as a baseline for comparison of the
more advanced schemes (HDF5, Zarr, and audio-based), one can conclude that even
plain bzip2 or LZMA compression can achieve comparable compression results. A
tradeoff to consider is the lack of metadata and internal structure, which might
cause additional data handling overhead as easy-to-use import and parsing tools
are not available. Variable-length encoding using LEB128S is a suitable input
for the bzip2 and LZMA compressors when combined with a column-based storage
format. Delta encoding resulted in comparably good CS in certain combinations.

Some datasets are inherently more compressible than others. This is a result
of the entropy analysis and can be observed in the data representation
evaluation as well. Compressing BLUED consistently yields smaller file sizes
with most compressors than any other dataset. The benchmark shows that higher
entropy correlates strongly with higher CS per dataset.

While the majority of tested data representation formats achieves a data
reduction, compared to the baseline, some formats are counter-productive and
generate a larger output (CS over 100\%). This behavior affects most HDF5- and
Zarr-based formats, because of the 1-byte overhead (depending on the used
compressor).

Choosing the best-performing data representation for each dataset, the following
SS can be achieved when applied to all data files as compared against the raw
binary encoding: \input{results/file_format_eval_summary} It can be noted that
REDD, UK-DALE, and both BLOND datasets perform at around 50-60\% of CS, while
BLUED shows a significantly smaller CS of below 30\% CS, due to it's very low
signal entropy (Table~\ref{tab:entropy-used-values}). Variable-length encoding
(LEB128S) and Delta encoding yield the largest space saving for such types of
data (REDD and BLUED).

Two out of the five evaluated datasets (REDD and BLUED) showed the highest space
savings with a general-purpose compressor (bzip2) and variable-length encoding.
ALAC and HDF5+MAFISC performed best on UK-DALE, BLOND-50, and BLOND-250, given
their higher signal entropy and value range utilization.

When comparing the raw space savings against the actually published dataset, which
typically is already compressed, we can achieve additional space savings:
\input{results/file_format_eval_summary_2} All datasets show space savings,
except for UK-DALE, which shows an insignificant increase in the overall dataset
size. This means the originally published FLAC files are already compressed to a
high extent; this is supported by Figure~\ref{fig:file-format-eval}, showing
FLAC among the highest ranking formats in this study. While an absolute space
saving of \SI{1.1}{\gibi\byte} for REDD might be insignificant in most use cases
(desktop computing and data center), a more compelling reduction in storage
space of up to \SI{1867.9}{\gibi\byte} for BLOND-250 can be substantially
beneficial.

\subsection{Chunk Size Impact}
\label{sec:results-chunk-size-impact}

\begin{figure}[!htbp]
  \centering
  \includegraphics[width=0.499\textwidth]{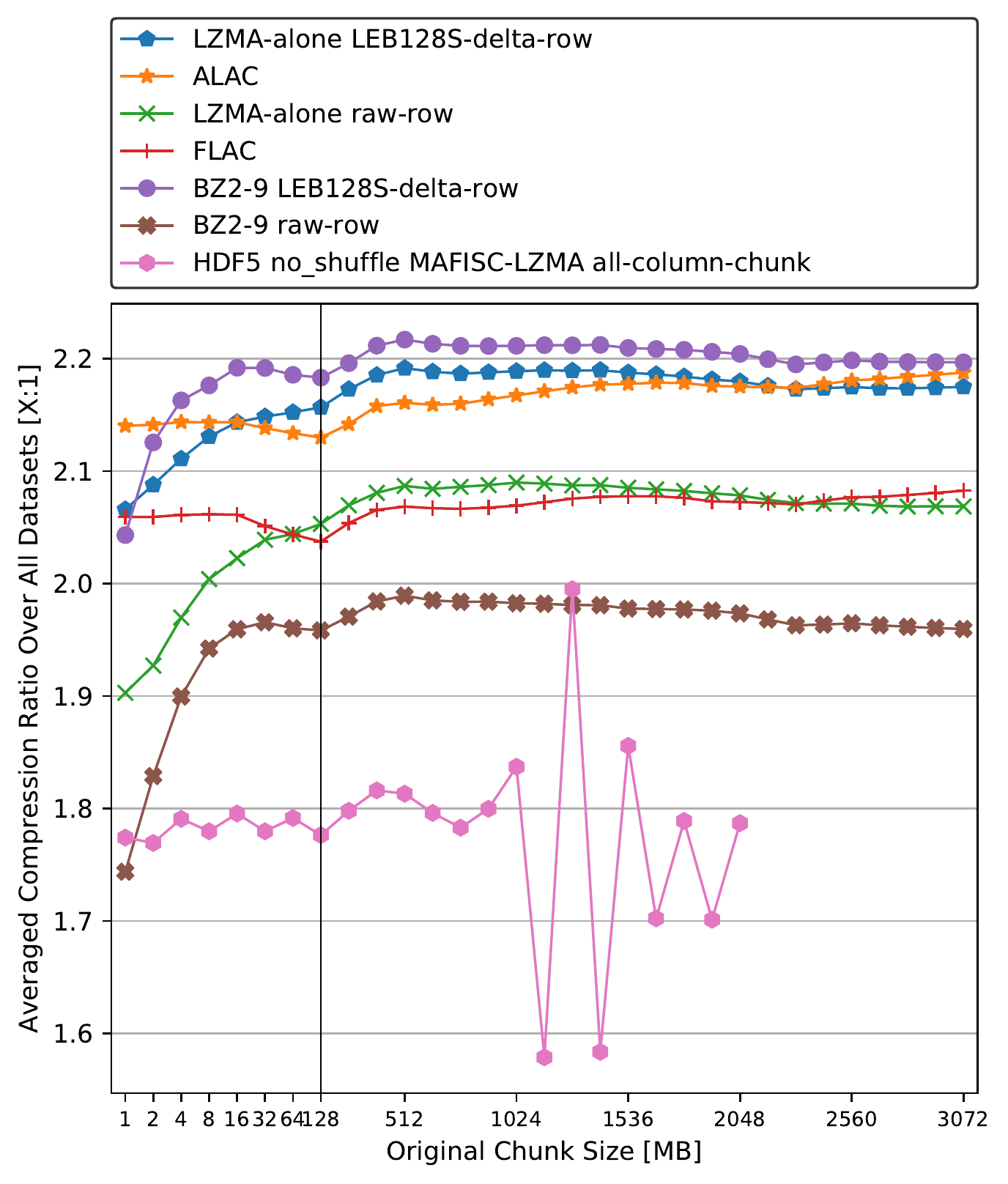}
  \caption{Chunk size impact of different representations.}
  \label{fig:chunk-size-impact}
\end{figure}

The chunk size evaluation (Figure~\ref{fig:chunk-size-impact}) contains  the
averaged CR per chunk size for all datasets except REDD, as it only contains
\SI{1438.4}{\mebi\byte} of data and was therefore omitted. A detailed
per-dataset evaluation is available in the online appendix\footnote{The online
appendix is available through the program chair (double-blind review).}.

The evaluated chunk size range starts with very small chunks, which would not be
recommended for large datasets because of the increased handling and container
overhead. As such, chunk sizes starting with \SI{128}{\mebi\byte} can be
considered as viable storage strategy. The resulting CR ramps up quickly for
most formats until it levels off between \SIrange{32}{64}{\mebi\byte}. Above
this mark, no significant improvement in CR can be achieved by increasing the
chunk size. Some file formats even show a slight linear decrease in CR with very
large chunk sizes (above approx. \SI{1.5}{\gibi\byte}). ALAC and FLAC
compressors show a slight improvement (2-3\%) in CR with larger chunk sizes. In
most use cases this size reduction comes at a great cost in RAM requirement to
process files above \SI{2048}{\mebi\byte}.  HDF5 has its own concept of
"chunks", used for I/O and the filter pipeline, with a default size of
\SI{1}{\mebi\byte}. Internal limitations do not allow for HDF5-chunks larger
than \SI{2048}{\mebi\byte}, however, HDF5, in general, can be used for files
larger than this limit. The MAFISC filter with LZMA compression experiences
large fluctuations for neighboring chunk size steps and should, therefore, be
tuned separately. Overall, increasing the chunk size has a negligible effect on
the final compression ratio and only pushes up the RAM requirements for
processing.

\subsection{Summary and Recommendations}
\label{sec:summary-and-recommendations}

The entropy analysis shows a lack of measurement range calibration in some
datasets. This results in unutilized precision, that would have been available
with the given hardware DAQ units. The used range directly affects the contained
entropy, and therefore the achievable compression ratio. A well-calibrated
measurement system is a key requirement to achieve the best signal range and
resolution.

Choosing a file format for long-term whole-building energy datasets is a crucial
component, directly affecting the visibility and accessibility of the data by
other researchers. Using an unsupported encoding or requiring specialized tools
to read the data is cumbersome and error-prone and should be avoided. We
recommend using well-known file formats, such as HDF5 or FLAC, which are widely
adopted and provide built-in support for metadata, compression, and
error-detection. While ALAC and FLAC already provide internal compression, we
recommend the MAFISC or zstd filters for HDF5, due to their superior compression
ratio. The serialization orientation (row- or column-based) has only a minor
effect.

Large datasets should be split into multiple smaller files to facilitate data
handling, reduce transfer speeds and loading times for short amounts of data. We
have found that compression algorithms (together with the above-described file
formats) yield higher space savings with chunk sizes above
\SIrange{256}{384}{\mebi\byte}. Small files show a modest compression ratio,
while larger files require more transfer bandwidth and time before the data can
be analyzed.

%% file: results/file_format_eval_summary.tex
REDD: 48.3\% or \SI{0.7}{\gibi\byte},
BLUED: 73.0\% or \SI{30.0}{\gibi\byte},
UK-DALE: 40.5\% or \SI{2534.1}{\gibi\byte},
BLOND-50: 51.3\% or \SI{5252.3}{\gibi\byte},
BLOND-250: 55.4\% or \SI{6590.8}{\gibi\byte}.

%% file: results/file_format_eval_summary_2.tex
REDD: 61.2\% or \SI{1.1}{\gibi\byte},
BLUED: 96.4\% or \SI{295.5}{\gibi\byte},
UK-DALE: -1.3\% or \SI{-49.1}{\gibi\byte},
BLOND-50: 23.3\% or \SI{1519.7}{\gibi\byte},
BLOND-250: 26.0\% or \SI{1867.9}{\gibi\byte}.

%% file: 08-conclusions.tex

\section{Conclusions}
\label{sec:conclusions}

We presented a comprehensive entropy analysis of public whole-building energy
datasets with waveform signals. Some datasets leave a majority of the available
ADC range unused, causing lost precision and accuracy. A well-calibrated
measurement system maximizes the achievable precision. Using
\numberOfDataRepresentations{} different data representation formats, we have
shown that immense space savings of up to 73\% are achievable by choosing a
suitable file format and data transformation. Low entropy datasets show higher
achievable compression ratios. Audio-based file formats perform considerably
well, given the similarities to electricity waveforms. Transparent data
transformations are particularly beneficial, such as MAFISC and SHUFFLE-based
approaches. The input size shows a mostly stable dependency to the achievable
compressed size, with variations of a few percentage points (limited by RAM).
Waveform data shows a nearly constant compression ratio, independent of the
input chunk size. Splitting large datasets into multiple smaller files is
important for data handling, but insignificant in terms of space savings.

\FloatBarrier

%% file: ms.bbl

\begin{thebibliography}{46}


\ifx \showCODEN    \undefined \def \showCODEN     #1{\unskip}     \fi
\ifx \showDOI      \undefined \def \showDOI       #1{#1}\fi
\ifx \showISBNx    \undefined \def \showISBNx     #1{\unskip}     \fi
\ifx \showISBNxiii \undefined \def \showISBNxiii  #1{\unskip}     \fi
\ifx \showISSN     \undefined \def \showISSN      #1{\unskip}     \fi
\ifx \showLCCN     \undefined \def \showLCCN      #1{\unskip}     \fi
\ifx \shownote     \undefined \def \shownote      #1{#1}          \fi
\ifx \showarticletitle \undefined \def \showarticletitle #1{#1}   \fi
\ifx \showURL      \undefined \def \showURL       {\relax}        \fi
\providecommand\bibfield[2]{#2}
\providecommand\bibinfo[2]{#2}
\providecommand\natexlab[1]{#1}
\providecommand\showeprint[2][]{arXiv:#2}

\bibitem[\protect\citeauthoryear{Alted}{Alted}{2017}]%
        {Blosc}
\bibfield{author}{\bibinfo{person}{Francesc Alted}.}
  \bibinfo{year}{2017}\natexlab{}.
\newblock \bibinfo{title}{{Blosc: A high performance compressor optimized for
  binary data}}.
\newblock   (\bibinfo{date}{November} \bibinfo{year}{2017}).
\newblock
\showURL{%
Retrieved January 20, 2018 from \url{http://blosc.org/}}


\bibitem[\protect\citeauthoryear{{American National Standards
  Institute}}{{American National Standards Institute}}{2016}]%
        {ANSI:C84.1}
\bibfield{author}{\bibinfo{person}{{American National Standards Institute}}.}
  \bibinfo{year}{2016}\natexlab{}.
\newblock \bibinfo{title}{{ANSI C84.1-2016: Standard for Electric Power Systems
  and Equipment—Voltage Ratings (60 Hz)}}.
\newblock   (\bibinfo{year}{2016}).
\newblock


\bibitem[\protect\citeauthoryear{Anderson, Ocneanu, Benitez, Carlson, Rowe, and
  Berges}{Anderson et~al\mbox{.}}{2012}]%
        {Anderson2012-BLUED}
\bibfield{author}{\bibinfo{person}{Kyle Anderson}, \bibinfo{person}{Adrian
  Ocneanu}, \bibinfo{person}{Diego Benitez}, \bibinfo{person}{Derrick Carlson},
  \bibinfo{person}{Anthony Rowe}, {and} \bibinfo{person}{Mario Berges}.}
  \bibinfo{year}{2012}\natexlab{}.
\newblock \showarticletitle{{BLUED: A Fully Labeled Public Dataset for
  Event-Based Non-Intrusive Load Monitoring Research}}. In
  \bibinfo{booktitle}{{\em SustKDD '12}}. \bibinfo{publisher}{ACM},
  \bibinfo{address}{Beijing, China}, \bibinfo{pages}{1--5}.
\newblock


\bibitem[\protect\citeauthoryear{Arnold and Bell}{Arnold and Bell}{1997}]%
        {Arnold1997}
\bibfield{author}{\bibinfo{person}{R. Arnold} {and} \bibinfo{person}{T. Bell}.}
  \bibinfo{year}{1997}\natexlab{}.
\newblock \showarticletitle{A corpus for the evaluation of lossless compression
  algorithms}. In \bibinfo{booktitle}{{\em Data Compression Conference, 1997.
  DCC '97. Proceedings}}. \bibinfo{pages}{201--210}.
\newblock
\showISSN{1068-0314}
\showDOI{%
\url{https://doi.org/10.1109/DCC.1997.582019}}


\bibitem[\protect\citeauthoryear{Association}{Association}{2018}]%
        {COMTRADE}
\bibfield{author}{\bibinfo{person}{IEEE~Standards Association}.}
  \bibinfo{year}{2018}\natexlab{}.
\newblock \bibinfo{title}{{COMTRADE: Common format for Transient Data Exchange
  for power systems}}.
\newblock   (\bibinfo{date}{January} \bibinfo{year}{2018}).
\newblock
\showURL{%
Retrieved January 20, 2018 from
  \url{https://standards.ieee.org/findstds/standard/C37.111-2013.html}}


\bibitem[\protect\citeauthoryear{Batra, Kelly, Parson, Dutta, Knottenbelt,
  Rogers, Singh, and Srivastava}{Batra et~al\mbox{.}}{2014}]%
        {NILMTK}
\bibfield{author}{\bibinfo{person}{Nipun Batra}, \bibinfo{person}{Jack Kelly},
  \bibinfo{person}{Oliver Parson}, \bibinfo{person}{Haimonti Dutta},
  \bibinfo{person}{William Knottenbelt}, \bibinfo{person}{Alex Rogers},
  \bibinfo{person}{Amarjeet Singh}, {and} \bibinfo{person}{Mani Srivastava}.}
  \bibinfo{year}{2014}\natexlab{}.
\newblock \showarticletitle{{NILMTK: An Open Source Toolkit for Non-intrusive
  Load Monitoring}}. In \bibinfo{booktitle}{{\em ACM e-Energy '14}}.
  \bibinfo{publisher}{ACM}, \bibinfo{address}{New York, NY, USA},
  \bibinfo{pages}{265--276}.
\newblock
\showDOI{%
\url{https://doi.org/10.1145/2602044.2602051}}


\bibitem[\protect\citeauthoryear{Blanas, Wu, Byna, Dong, and Shoshani}{Blanas
  et~al\mbox{.}}{2014}]%
        {Blanas2014}
\bibfield{author}{\bibinfo{person}{Spyros Blanas}, \bibinfo{person}{Kesheng
  Wu}, \bibinfo{person}{Surendra Byna}, \bibinfo{person}{Bin Dong}, {and}
  \bibinfo{person}{Arie Shoshani}.} \bibinfo{year}{2014}\natexlab{}.
\newblock \showarticletitle{Parallel Data Analysis Directly on Scientific File
  Formats}. In \bibinfo{booktitle}{{\em Proceedings of the 2014 ACM SIGMOD
  International Conference on Management of Data}} {\em
  (\bibinfo{series}{SIGMOD '14})}. \bibinfo{publisher}{ACM},
  \bibinfo{address}{New York, NY, USA}, \bibinfo{pages}{385--396}.
\newblock
\showISBNx{978-1-4503-2376-5}
\showDOI{%
\url{https://doi.org/10.1145/2588555.2612185}}


\bibitem[\protect\citeauthoryear{Bookstein and Storer}{Bookstein and
  Storer}{1992}]%
        {Bookstein1992}
\bibfield{author}{\bibinfo{person}{Abraham Bookstein} {and}
  \bibinfo{person}{James~A. Storer}.} \bibinfo{year}{1992}\natexlab{}.
\newblock \showarticletitle{Data compression}.
\newblock \bibinfo{journal}{{\em Information Processing \& Management\/}}
  \bibinfo{volume}{28}, \bibinfo{number}{6} (\bibinfo{year}{1992}),
  \bibinfo{pages}{675 -- 680}.
\newblock
\showISSN{0306-4573}
\showDOI{%
\url{https://doi.org/10.1016/0306-4573(92)90060-D}}
\newblock
\shownote{Special Issue: Data compression for images and texts.}


\bibitem[\protect\citeauthoryear{Bryant}{Bryant}{2018}]%
        {WavPack}
\bibfield{author}{\bibinfo{person}{David Bryant}.}
  \bibinfo{year}{2018}\natexlab{}.
\newblock \bibinfo{title}{{WavPack: Hybrid Lossless Audio Compression}}.
\newblock   (\bibinfo{date}{January} \bibinfo{year}{2018}).
\newblock
\showURL{%
Retrieved January 20, 2018 from \url{http://www.wavpack.com/}}


\bibitem[\protect\citeauthoryear{de~Souza, Assis, and Pal}{de~Souza
  et~al\mbox{.}}{2017}]%
        {deSouza2017}
\bibfield{author}{\bibinfo{person}{J.~C.~S. de Souza},
  \bibinfo{person}{T.~M.~L. Assis}, {and} \bibinfo{person}{B.~C. Pal}.}
  \bibinfo{year}{2017}\natexlab{}.
\newblock \showarticletitle{Data Compression in Smart Distribution Systems via
  Singular Value Decomposition}.
\newblock \bibinfo{journal}{{\em IEEE Transactions on Smart Grid\/}}
  \bibinfo{volume}{8}, \bibinfo{number}{1} (\bibinfo{date}{Jan}
  \bibinfo{year}{2017}), \bibinfo{pages}{275--284}.
\newblock
\showISSN{1949-3053}
\showDOI{%
\url{https://doi.org/10.1109/TSG.2015.2456979}}


\bibitem[\protect\citeauthoryear{Deelman and Chervenak}{Deelman and
  Chervenak}{2008}]%
        {Deelman2008}
\bibfield{author}{\bibinfo{person}{E. Deelman} {and} \bibinfo{person}{A.
  Chervenak}.} \bibinfo{year}{2008}\natexlab{}.
\newblock \showarticletitle{Data Management Challenges of Data-Intensive
  Scientific Workflows}. In \bibinfo{booktitle}{{\em 2008 Eighth IEEE
  International Symposium on Cluster Computing and the Grid (CCGRID)}}.
  \bibinfo{pages}{687--692}.
\newblock
\showDOI{%
\url{https://doi.org/10.1109/CCGRID.2008.24}}


\bibitem[\protect\citeauthoryear{Dougherty, Folk, Zadok, Bernstein, Bernstein,
  Eliceiri, Benger, and Best}{Dougherty et~al\mbox{.}}{2009}]%
        {Dougherty2009}
\bibfield{author}{\bibinfo{person}{Matthew~T. Dougherty},
  \bibinfo{person}{Michael~J. Folk}, \bibinfo{person}{Erez Zadok},
  \bibinfo{person}{Herbert~J. Bernstein}, \bibinfo{person}{Frances~C.
  Bernstein}, \bibinfo{person}{Kevin~W. Eliceiri}, \bibinfo{person}{Werner
  Benger}, {and} \bibinfo{person}{Christoph Best}.}
  \bibinfo{year}{2009}\natexlab{}.
\newblock \showarticletitle{Unifying Biological Image Formats with HDF5}.
\newblock \bibinfo{journal}{{\em Commun. ACM\/}} \bibinfo{volume}{52},
  \bibinfo{number}{10} (\bibinfo{date}{Oct.} \bibinfo{year}{2009}),
  \bibinfo{pages}{42--47}.
\newblock
\showISSN{0001-0782}
\showDOI{%
\url{https://doi.org/10.1145/1562764.1562781}}


\bibitem[\protect\citeauthoryear{Eichinger, Efros, Karnouskos, and
  B\"{o}hm}{Eichinger et~al\mbox{.}}{2015}]%
        {Eichinger2015}
\bibfield{author}{\bibinfo{person}{Frank Eichinger}, \bibinfo{person}{Pavel
  Efros}, \bibinfo{person}{Stamatis Karnouskos}, {and} \bibinfo{person}{Klemens
  B\"{o}hm}.} \bibinfo{year}{2015}\natexlab{}.
\newblock \showarticletitle{A Time-series Compression Technique and Its
  Application to the Smart Grid}.
\newblock \bibinfo{journal}{{\em The VLDB Journal\/}} \bibinfo{volume}{24},
  \bibinfo{number}{2} (\bibinfo{date}{April} \bibinfo{year}{2015}),
  \bibinfo{pages}{193--218}.
\newblock
\showISSN{1066-8888}
\showDOI{%
\url{https://doi.org/10.1007/s00778-014-0368-8}}


\bibitem[\protect\citeauthoryear{{European Committee for Electrotechnical
  Standardization}}{{European Committee for Electrotechnical
  Standardization}}{1989}]%
        {CENELEC:HD472S1}
\bibfield{author}{\bibinfo{person}{{European Committee for Electrotechnical
  Standardization}}.} \bibinfo{year}{1989}\natexlab{}.
\newblock \bibinfo{title}{{CENELEC Harmonisation Document HD 472 S1}}.
\newblock   (\bibinfo{year}{1989}).
\newblock


\bibitem[\protect\citeauthoryear{Folk, Heber, Koziol, Pourmal, and
  Robinson}{Folk et~al\mbox{.}}{2011}]%
        {Folk2011}
\bibfield{author}{\bibinfo{person}{Mike Folk}, \bibinfo{person}{Gerd Heber},
  \bibinfo{person}{Quincey Koziol}, \bibinfo{person}{Elena Pourmal}, {and}
  \bibinfo{person}{Dana Robinson}.} \bibinfo{year}{2011}\natexlab{}.
\newblock \showarticletitle{An Overview of the HDF5 Technology Suite and Its
  Applications}. In \bibinfo{booktitle}{{\em Proceedings of the EDBT/ICDT 2011
  Workshop on Array Databases}} {\em (\bibinfo{series}{AD '11})}.
  \bibinfo{publisher}{ACM}, \bibinfo{address}{New York, NY, USA},
  \bibinfo{pages}{36--47}.
\newblock
\showISBNx{978-1-4503-0614-0}
\showDOI{%
\url{https://doi.org/10.1145/1966895.1966900}}


\bibitem[\protect\citeauthoryear{Foundation}{Foundation}{2018}]%
        {FLAC}
\bibfield{author}{\bibinfo{person}{Xiph.Org Foundation}.}
  \bibinfo{year}{2018}\natexlab{}.
\newblock \bibinfo{title}{{FLAC: Free Lossless Audio Codec}}.
\newblock   (\bibinfo{date}{January} \bibinfo{year}{2018}).
\newblock
\showURL{%
Retrieved January 20, 2018 from \url{https://xiph.org/flac/}}


\bibitem[\protect\citeauthoryear{Gerek and Ece}{Gerek and Ece}{2004}]%
        {Gerek2004}
\bibfield{author}{\bibinfo{person}{O.~N. Gerek} {and} \bibinfo{person}{D.~G.
  Ece}.} \bibinfo{year}{2004}\natexlab{}.
\newblock \showarticletitle{2-D analysis and compression of power-quality event
  data}.
\newblock \bibinfo{journal}{{\em IEEE Transactions on Power Delivery\/}}
  \bibinfo{volume}{19}, \bibinfo{number}{2} (\bibinfo{date}{April}
  \bibinfo{year}{2004}), \bibinfo{pages}{791--798}.
\newblock
\showISSN{0885-8977}
\showDOI{%
\url{https://doi.org/10.1109/TPWRD.2003.823197}}


\bibitem[\protect\citeauthoryear{Gosink, Shalf, Stockinger, Wu, and
  Bethel}{Gosink et~al\mbox{.}}{2006}]%
        {Gosink2006}
\bibfield{author}{\bibinfo{person}{L. Gosink}, \bibinfo{person}{J. Shalf},
  \bibinfo{person}{K. Stockinger}, \bibinfo{person}{Kesheng Wu}, {and}
  \bibinfo{person}{W. Bethel}.} \bibinfo{year}{2006}\natexlab{}.
\newblock \showarticletitle{HDF5-FastQuery: Accelerating Complex Queries on HDF
  Datasets using Fast Bitmap Indices}. In \bibinfo{booktitle}{{\em 18th
  International Conference on Scientific and Statistical Database Management
  (SSDBM'06)}}. \bibinfo{pages}{149--158}.
\newblock
\showISSN{1551-6393}
\showDOI{%
\url{https://doi.org/10.1109/SSDBM.2006.27}}


\bibitem[\protect\citeauthoryear{Group}{Group}{2018}]%
        {DWARF3.0}
\bibfield{author}{\bibinfo{person}{Free~Standards Group}.}
  \bibinfo{year}{2018}\natexlab{}.
\newblock \bibinfo{title}{{DWARF Debugging Information Format Specification
  Version 3.0}}.
\newblock   (\bibinfo{date}{January} \bibinfo{year}{2018}).
\newblock
\showURL{%
Retrieved January 20, 2018 from \url{http://dwarfstd.org/doc/Dwarf3.pdf}}


\bibitem[\protect\citeauthoryear{Group}{Group}{2017}]%
        {szip}
\bibfield{author}{\bibinfo{person}{HDF Group}.}
  \bibinfo{year}{2017}\natexlab{}.
\newblock \bibinfo{title}{{Szip Compression in HDF Products}}.
\newblock   (\bibinfo{date}{November} \bibinfo{year}{2017}).
\newblock
\showURL{%
Retrieved January 20, 2018 from
  \url{https://support.hdfgroup.org/doc_resource/SZIP/}}


\bibitem[\protect\citeauthoryear{Haq, Kriechbaumer, Kahl, and Jacobsen}{Haq
  et~al\mbox{.}}{2017}]%
        {Haq2017-CLEAR}
\bibfield{author}{\bibinfo{person}{Anwar~Ul Haq}, \bibinfo{person}{Thomas
  Kriechbaumer}, \bibinfo{person}{Matthias Kahl}, {and}
  \bibinfo{person}{Hans-Arno Jacobsen}.} \bibinfo{year}{2017}\natexlab{}.
\newblock \showarticletitle{{CLEAR – A Circuit Level Electric Appliance Radar
  for the Electric Cabinet}}. In \bibinfo{booktitle}{{\em 2017 IEEE
  International Conference on Industrial Technology}} {\em
  (\bibinfo{series}{ICIT '17})}. \bibinfo{pages}{1130--1135}.
\newblock
\showISBNx{978-1-5090-5319-3/17}
\showDOI{%
\url{https://doi.org/10.1109/ICIT.2017.7915521}}


\bibitem[\protect\citeauthoryear{H{\"u}bbe and Kunkel}{H{\"u}bbe and
  Kunkel}{2013}]%
        {Huebbe2013}
\bibfield{author}{\bibinfo{person}{Nathanael H{\"u}bbe} {and}
  \bibinfo{person}{Julian Kunkel}.} \bibinfo{year}{2013}\natexlab{}.
\newblock \showarticletitle{Reducing the HPC-datastorage footprint with
  MAFISC---Multidimensional Adaptive Filtering Improved Scientific data
  Compression}.
\newblock \bibinfo{journal}{{\em Computer Science - Research and
  Development\/}} \bibinfo{volume}{28}, \bibinfo{number}{2} (\bibinfo{date}{01
  May} \bibinfo{year}{2013}), \bibinfo{pages}{231--239}.
\newblock
\showISSN{1865-2042}
\showDOI{%
\url{https://doi.org/10.1007/s00450-012-0222-4}}


\bibitem[\protect\citeauthoryear{Inc.}{Inc.}{2018}]%
        {ALAC}
\bibfield{author}{\bibinfo{person}{Apple Inc.}}
  \bibinfo{year}{2018}\natexlab{}.
\newblock \bibinfo{title}{{ALAC: Apple Lossless Audio Codec}}.
\newblock   (\bibinfo{date}{January} \bibinfo{year}{2018}).
\newblock
\showURL{%
Retrieved January 20, 2018 from \url{https://macosforge.github.io/alac/}}


\bibitem[\protect\citeauthoryear{Kahl, Haq, Kriechbaumer, and Jacobsen}{Kahl
  et~al\mbox{.}}{2017}]%
        {Kahl2017}
\bibfield{author}{\bibinfo{person}{Matthias Kahl}, \bibinfo{person}{Anwar~Ul
  Haq}, \bibinfo{person}{Thomas Kriechbaumer}, {and} \bibinfo{person}{Hans-Arno
  Jacobsen}.} \bibinfo{year}{2017}\natexlab{}.
\newblock \showarticletitle{{A Comprehensive Feature Study for Appliance
  Recognition on High Frequency Energy Data}}. In \bibinfo{booktitle}{{\em
  Proceedings of the 2017 ACM Eighth International Conference on Future Energy
  Systems}} {\em (\bibinfo{series}{e-Energy '17})}. \bibinfo{publisher}{ACM},
  \bibinfo{address}{New York, NY, USA}.
\newblock
\showISBNx{978-1-4503-5036-5/17/05}
\showDOI{%
\url{https://doi.org/10.1145/3077839.3077845}}


\bibitem[\protect\citeauthoryear{Kelly and Knottenbelt}{Kelly and
  Knottenbelt}{2015}]%
        {Kelly2015-UK-DALE}
\bibfield{author}{\bibinfo{person}{Jack Kelly} {and} \bibinfo{person}{William
  Knottenbelt}.} \bibinfo{year}{2015}\natexlab{}.
\newblock \showarticletitle{{The UK-DALE dataset, domestic appliance-level
  electricity demand and whole-house demand from five UK homes}}.
\newblock \bibinfo{journal}{{\em Scientific Data\/}} \bibinfo{volume}{2},
  \bibinfo{number}{150007} (\bibinfo{year}{2015}).
\newblock
\showDOI{%
\url{https://doi.org/10.1038/sdata.2015.7}}


\bibitem[\protect\citeauthoryear{Kolter and Johnson}{Kolter and Johnson}{[n.
  d.]}]%
        {Kolter2011-REDD}
\bibfield{author}{\bibinfo{person}{J.~Zico Kolter} {and}
  \bibinfo{person}{Matthew~J. Johnson}.} \bibinfo{year}{[n. d.]}\natexlab{}.
\newblock \showarticletitle{{REDD: A Public Data Set for Energy Disaggregation
  Research}}. In \bibinfo{booktitle}{{\em SustKDD '11}} (2011),
  Vol.~\bibinfo{volume}{25}. \bibinfo{pages}{59--62}.
\newblock


\bibitem[\protect\citeauthoryear{Kriechbaumer, Haq, Kahl, and
  Jacobsen}{Kriechbaumer et~al\mbox{.}}{2017}]%
        {Kriechbaumer2017-MEDAL}
\bibfield{author}{\bibinfo{person}{Thomas Kriechbaumer},
  \bibinfo{person}{Anwar~Ul Haq}, \bibinfo{person}{Matthias Kahl}, {and}
  \bibinfo{person}{Hans-Arno Jacobsen}.} \bibinfo{year}{2017}\natexlab{}.
\newblock \showarticletitle{{MEDAL: A Cost-Effective High-Frequency Energy Data
  Acquisition System for Electrical Appliances}}. In \bibinfo{booktitle}{{\em
  Proceedings of the 2017 ACM Eighth International Conference on Future Energy
  Systems}} {\em (\bibinfo{series}{e-Energy '17})}. \bibinfo{publisher}{ACM},
  \bibinfo{address}{New York, NY, USA}.
\newblock
\showISBNx{978-1-4503-5036-5/17/05}
\showDOI{%
\url{https://doi.org/10.1145/3077839.3077844}}


\bibitem[\protect\citeauthoryear{Kriechbaumer and Jacobsen}{Kriechbaumer and
  Jacobsen}{2018}]%
        {Kriechbaumer2018-BLOND}
\bibfield{author}{\bibinfo{person}{Thomas Kriechbaumer} {and}
  \bibinfo{person}{Hans-Arno Jacobsen}.} \bibinfo{year}{2018}\natexlab{}.
\newblock \bibinfo{title}{{BLOND, a building-level office environment dataset
  of typical electrical appliances}}.
\newblock   (\bibinfo{date}{March} \bibinfo{year}{2018}).
\newblock
\showDOI{%
\url{https://doi.org/10.1038/sdata.2018.48}}


\bibitem[\protect\citeauthoryear{Liu and Shen}{Liu and Shen}{2017}]%
        {Liu2017}
\bibfield{author}{\bibinfo{person}{Guoxin Liu} {and} \bibinfo{person}{Haiying
  Shen}.} \bibinfo{year}{2017}\natexlab{}.
\newblock \showarticletitle{Minimum-Cost Cloud Storage Service Across Multiple
  Cloud Providers}.
\newblock \bibinfo{journal}{{\em IEEE/ACM Trans. Netw.\/}}
  \bibinfo{volume}{25}, \bibinfo{number}{4} (\bibinfo{date}{Aug.}
  \bibinfo{year}{2017}), \bibinfo{pages}{2498--2513}.
\newblock
\showISSN{1063-6692}
\showDOI{%
\url{https://doi.org/10.1109/TNET.2017.2693222}}


\bibitem[\protect\citeauthoryear{Masui, Amiri, Connor, Deng, Fandino,
  H{\"o}fer, Halpern, Hanna, Hincks, Hinshaw, Parra, Newburgh, Shaw, and
  Vanderlinde}{Masui et~al\mbox{.}}{2015}]%
        {Masui2015}
\bibfield{author}{\bibinfo{person}{K. Masui}, \bibinfo{person}{M. Amiri},
  \bibinfo{person}{L. Connor}, \bibinfo{person}{M. Deng}, \bibinfo{person}{M.
  Fandino}, \bibinfo{person}{C. H{\"o}fer}, \bibinfo{person}{M. Halpern},
  \bibinfo{person}{D. Hanna}, \bibinfo{person}{A.D. Hincks},
  \bibinfo{person}{G. Hinshaw}, \bibinfo{person}{J.M. Parra},
  \bibinfo{person}{L.B. Newburgh}, \bibinfo{person}{J.R. Shaw}, {and}
  \bibinfo{person}{K. Vanderlinde}.} \bibinfo{year}{2015}\natexlab{}.
\newblock \showarticletitle{A compression scheme for radio data in high
  performance computing}.
\newblock \bibinfo{journal}{{\em Astronomy and Computing\/}}
  \bibinfo{volume}{12}, \bibinfo{number}{Supplement C} (\bibinfo{year}{2015}),
  \bibinfo{pages}{181 -- 190}.
\newblock
\showISSN{2213-1337}
\showDOI{%
\url{https://doi.org/10.1016/j.ascom.2015.07.002}}


\bibitem[\protect\citeauthoryear{Meziane, Picon, Ravier, Lamarque, Bunetel, and
  Raingeaud}{Meziane et~al\mbox{.}}{2016}]%
        {Meziane2016}
\bibfield{author}{\bibinfo{person}{M.~N. Meziane}, \bibinfo{person}{T. Picon},
  \bibinfo{person}{P. Ravier}, \bibinfo{person}{G. Lamarque},
  \bibinfo{person}{J.~C.~Le Bunetel}, {and} \bibinfo{person}{Y. Raingeaud}.}
  \bibinfo{year}{2016}\natexlab{}.
\newblock \showarticletitle{{A Measurement System for Creating Datasets of
  On/Off-Controlled Electrical Loads}}. In \bibinfo{booktitle}{{\em 2016 IEEE
  16th International Conference on Environment and Electrical Engineering
  (EEEIC)}}. \bibinfo{pages}{1--5}.
\newblock
\showDOI{%
\url{https://doi.org/10.1109/EEEIC.2016.7555847}}


\bibitem[\protect\citeauthoryear{Miles}{Miles}{2018}]%
        {Zarr}
\bibfield{author}{\bibinfo{person}{Alistair Miles}.}
  \bibinfo{year}{2018}\natexlab{}.
\newblock \bibinfo{title}{{Zarr: A Python package providing an implementation
  of chunked, compressed, N-dimensional arrays}}.
\newblock   (\bibinfo{date}{January} \bibinfo{year}{2018}).
\newblock
\showURL{%
Retrieved January 20, 2018 from \url{https://zarr.readthedocs.io/en/latest/}}


\bibitem[\protect\citeauthoryear{Nabeel, Javed, and Arshad}{Nabeel
  et~al\mbox{.}}{2013}]%
        {Nabeel2013}
\bibfield{author}{\bibinfo{person}{Muhammad Nabeel}, \bibinfo{person}{Fahad
  Javed}, {and} \bibinfo{person}{Naveed Arshad}.}
  \bibinfo{year}{2013}\natexlab{}.
\newblock \showarticletitle{Towards Smart Data Compression for Future Energy
  Management System}. In \bibinfo{booktitle}{{\em Fifth International
  Conference on Applied Energy}}.
\newblock


\bibitem[\protect\citeauthoryear{Paris, Donnal, and Leeb}{Paris
  et~al\mbox{.}}{2014}]%
        {Paris2014}
\bibfield{author}{\bibinfo{person}{J. Paris}, \bibinfo{person}{J.~S. Donnal},
  {and} \bibinfo{person}{S.~B. Leeb}.} \bibinfo{year}{2014}\natexlab{}.
\newblock \showarticletitle{NilmDB: The Non-Intrusive Load Monitor Database}.
\newblock \bibinfo{journal}{{\em IEEE Transactions on Smart Grid\/}}
  \bibinfo{volume}{5}, \bibinfo{number}{5} (\bibinfo{date}{Sept}
  \bibinfo{year}{2014}), \bibinfo{pages}{2459--2467}.
\newblock
\showISSN{1949-3053}
\showDOI{%
\url{https://doi.org/10.1109/TSG.2014.2321582}}


\bibitem[\protect\citeauthoryear{Pereira}{Pereira}{2017}]%
        {Pereira2017-EMD-DF}
\bibfield{author}{\bibinfo{person}{Lucas Pereira}.}
  \bibinfo{year}{2017}\natexlab{}.
\newblock \showarticletitle{EMD-DF: A Data Model and File Format for Energy
  Disaggregation Datasets}. In \bibinfo{booktitle}{{\em Proceedings of the 4th
  ACM International Conference on Systems for Energy-Efficient Built
  Environments}} {\em (\bibinfo{series}{BuildSys '17})}.
  \bibinfo{publisher}{ACM}, \bibinfo{address}{New York, NY, USA}, Article
  \bibinfo{articleno}{52}, \bibinfo{numpages}{2}~pages.
\newblock
\showISBNx{978-1-4503-5544-5}
\showDOI{%
\url{https://doi.org/10.1145/3137133.3141474}}


\bibitem[\protect\citeauthoryear{Pereira, Nunes, and Berg{\'e}s}{Pereira
  et~al\mbox{.}}{2014}]%
        {Pereira2014}
\bibfield{author}{\bibinfo{person}{Lucas Pereira}, \bibinfo{person}{Nuno
  Nunes}, {and} \bibinfo{person}{Mario Berg{\'e}s}.}
  \bibinfo{year}{2014}\natexlab{}.
\newblock \showarticletitle{SURF and SURF-PI: A File Format and API for
  Non-intrusive Load Monitoring Public Datasets}. In \bibinfo{booktitle}{{\em
  Proceedings of the 5th International Conference on Future Energy Systems}}
  {\em (\bibinfo{series}{e-Energy '14})}. \bibinfo{publisher}{ACM},
  \bibinfo{address}{New York, NY, USA}, \bibinfo{pages}{225--226}.
\newblock
\showISBNx{978-1-4503-2819-7}
\showDOI{%
\url{https://doi.org/10.1145/2602044.2602078}}


\bibitem[\protect\citeauthoryear{Puttaswamy, Nandagopal, and
  Kodialam}{Puttaswamy et~al\mbox{.}}{2012}]%
        {Puttaswamy2012}
\bibfield{author}{\bibinfo{person}{Krishna~P.N. Puttaswamy},
  \bibinfo{person}{Thyaga Nandagopal}, {and} \bibinfo{person}{Murali
  Kodialam}.} \bibinfo{year}{2012}\natexlab{}.
\newblock \showarticletitle{Frugal Storage for Cloud File Systems}. In
  \bibinfo{booktitle}{{\em Proceedings of the 7th ACM European Conference on
  Computer Systems}} {\em (\bibinfo{series}{EuroSys '12})}.
  \bibinfo{publisher}{ACM}, \bibinfo{address}{New York, NY, USA},
  \bibinfo{pages}{71--84}.
\newblock
\showISBNx{978-1-4503-1223-3}
\showDOI{%
\url{https://doi.org/10.1145/2168836.2168845}}


\bibitem[\protect\citeauthoryear{Qing, Hongtao, Zhikun, and Zhiwen}{Qing
  et~al\mbox{.}}{2011}]%
        {Qing2011}
\bibfield{author}{\bibinfo{person}{A. Qing}, \bibinfo{person}{Z. Hongtao},
  \bibinfo{person}{H. Zhikun}, {and} \bibinfo{person}{C. Zhiwen}.}
  \bibinfo{year}{2011}\natexlab{}.
\newblock \showarticletitle{A Compression Approach of Power Quality Monitoring
  Data Based on Two-dimension DCT}. In \bibinfo{booktitle}{{\em 2011 Third
  International Conference on Measuring Technology and Mechatronics
  Automation}}, Vol.~\bibinfo{volume}{1}. \bibinfo{pages}{20--24}.
\newblock
\showISSN{2157-1473}
\showDOI{%
\url{https://doi.org/10.1109/ICMTMA.2011.12}}


\bibitem[\protect\citeauthoryear{Ringwelski, Renner, Reinhardt, Weigel, and
  Turau}{Ringwelski et~al\mbox{.}}{2012}]%
        {Ringwelski2012}
\bibfield{author}{\bibinfo{person}{Martin Ringwelski},
  \bibinfo{person}{Christian Renner}, \bibinfo{person}{Andreas Reinhardt},
  \bibinfo{person}{Andreas Weigel}, {and} \bibinfo{person}{Volker Turau}.}
  \bibinfo{year}{2012}\natexlab{}.
\newblock \showarticletitle{The Hitchhiker's Guide to choosing the Compression
  Algorithm for your Smart Meter Data}.
\newblock  (\bibinfo{date}{September} \bibinfo{year}{2012}),
  \bibinfo{pages}{935--940}.
\newblock
\showDOI{%
\url{https://doi.org/10.1109/EnergyCon.2012.6348285}}


\bibitem[\protect\citeauthoryear{Sehrish, Kowalkowski, Paterno, and
  Green}{Sehrish et~al\mbox{.}}{2017}]%
        {Sehrish2017}
\bibfield{author}{\bibinfo{person}{S. Sehrish}, \bibinfo{person}{J.
  Kowalkowski}, \bibinfo{person}{M. Paterno}, {and} \bibinfo{person}{C.
  Green}.} \bibinfo{year}{2017}\natexlab{}.
\newblock \showarticletitle{Python and HPC for High Energy Physics Data
  Analyses}. In \bibinfo{booktitle}{{\em Proceedings of the 7th Workshop on
  Python for High-Performance and Scientific Computing}} {\em
  (\bibinfo{series}{PyHPC'17})}. \bibinfo{publisher}{ACM},
  \bibinfo{address}{New York, NY, USA}, Article \bibinfo{articleno}{8},
  \bibinfo{numpages}{8}~pages.
\newblock
\showISBNx{978-1-4503-5124-9}
\showDOI{%
\url{https://doi.org/10.1145/3149869.3149877}}


\bibitem[\protect\citeauthoryear{Shannon}{Shannon}{1949}]%
        {Shannon1949}
\bibfield{author}{\bibinfo{person}{C.~E. Shannon}.}
  \bibinfo{year}{1949}\natexlab{}.
\newblock \showarticletitle{{Communication in the Presence of Noise}}.
\newblock \bibinfo{journal}{{\em Proceedings of the IRE\/}}
  \bibinfo{volume}{37}, \bibinfo{number}{1} (\bibinfo{date}{Jan}
  \bibinfo{year}{1949}), \bibinfo{pages}{10--21}.
\newblock
\showISSN{0096-8390}
\showDOI{%
\url{https://doi.org/10.1109/JRPROC.1949.232969}}


\bibitem[\protect\citeauthoryear{Society}{Society}{2018}]%
        {PQDIF}
\bibfield{author}{\bibinfo{person}{IEEE Power \&~Energy Society}.}
  \bibinfo{year}{2018}\natexlab{}.
\newblock \bibinfo{title}{{IEEE 1159 - PQDIF: Power Quality and Quantity Data
  Interchange Format}}.
\newblock   (\bibinfo{date}{January} \bibinfo{year}{2018}).
\newblock
\showURL{%
Retrieved January 20, 2018 from
  \url{http://grouper.ieee.org/groups/1159/3/docs.html}}


\bibitem[\protect\citeauthoryear{Tariq, Arshad, and Nabeel}{Tariq
  et~al\mbox{.}}{2015}]%
        {Tariq2015}
\bibfield{author}{\bibinfo{person}{Z.~B. Tariq}, \bibinfo{person}{N. Arshad},
  {and} \bibinfo{person}{M. Nabeel}.} \bibinfo{year}{2015}\natexlab{}.
\newblock \showarticletitle{Enhanced LZMA and BZIP2 for improved energy data
  compression}. In \bibinfo{booktitle}{{\em 2015 International Conference on
  Smart Cities and Green ICT Systems (SMARTGREENS)}}. \bibinfo{pages}{1--8}.
\newblock


\bibitem[\protect\citeauthoryear{Unterweger and Engel}{Unterweger and
  Engel}{2015}]%
        {Unterweger2015a}
\bibfield{author}{\bibinfo{person}{Andreas Unterweger} {and}
  \bibinfo{person}{Dominik Engel}.} \bibinfo{year}{2015}\natexlab{}.
\newblock \showarticletitle{Resumable load data compression in smart grids}.
\newblock \bibinfo{journal}{{\em IEEE Transactions on Smart Grid\/}}
  \bibinfo{volume}{6}, \bibinfo{number}{2} (\bibinfo{year}{2015}),
  \bibinfo{pages}{919--929}.
\newblock
\showDOI{%
\url{https://doi.org/10.1109/TSG.2014.2364686}}


\bibitem[\protect\citeauthoryear{Unterweger, Engel, and Ringwelski}{Unterweger
  et~al\mbox{.}}{2015}]%
        {Unterweger2015b}
\bibfield{author}{\bibinfo{person}{Andreas Unterweger},
  \bibinfo{person}{Dominik Engel}, {and} \bibinfo{person}{Martin Ringwelski}.}
  \bibinfo{year}{2015}\natexlab{}.
\newblock \bibinfo{booktitle}{{\em The Effect of Data Granularity on Load Data
  Compression}}.
\newblock \bibinfo{publisher}{Springer International Publishing},
  \bibinfo{address}{Cham}, \bibinfo{pages}{69--80}.
\newblock
\showISBNx{978-3-319-25876-8}
\showDOI{%
\url{https://doi.org/10.1007/978-3-319-25876-8_7}}


\bibitem[\protect\citeauthoryear{Yuan, Yang, Liu, and Chen}{Yuan
  et~al\mbox{.}}{2010}]%
        {Yuan2010}
\bibfield{author}{\bibinfo{person}{D. Yuan}, \bibinfo{person}{Y. Yang},
  \bibinfo{person}{X. Liu}, {and} \bibinfo{person}{J. Chen}.}
  \bibinfo{year}{2010}\natexlab{}.
\newblock \showarticletitle{A cost-effective strategy for intermediate data
  storage in scientific cloud workflow systems}. In \bibinfo{booktitle}{{\em
  2010 IEEE International Symposium on Parallel Distributed Processing
  (IPDPS)}}. \bibinfo{pages}{1--12}.
\newblock
\showISSN{1530-2075}
\showDOI{%
\url{https://doi.org/10.1109/IPDPS.2010.5470453}}


\end{thebibliography}
